\documentclass[10pt]{article}
\usepackage{amsfonts}
\usepackage{amsmath}
\usepackage{amssymb}
\usepackage{graphicx}
\def \be {\begin{equation}}
\def \ee {\end{equation}}
\def \bea {\begin{eqnarray}}
\def \eea {\end{eqnarray}}
\def \nn {\nonumber}

\def \rr {\raise.35ex\hbox{\small $\prime$}\kern-.17em{\mbox{\large $\imath$}}}
\def \del {\partial}
\def \dels {\partial\kern-.5em / \kern.5em}
\def \As {{A\kern-.5em / \kern.5em}}
\def \Ds {D\kern-.7em / \kern.5em}

\def \a {\alpha}
\def \ap {\alpha}

\def \lam {\lambda}

\def \Z {{\mathbb{Z}}}

\newenvironment{proof}[1][Proof]{\noindent\textbf{#1.} }{ \rule{0.5em}{0.5em}}
\def \cT {{\cal T}}

\def \tL {\tilde{L}}

\begin{document}
\begin{titlepage}
\begin{center}
\hfill hep-th/0504138\\
\vskip .6in
\textbf{\LARGE
Solving all 4-point correlation functions
for bosonic open string theory
in the high energy limit}
\vskip .5in
{\large Chuan-Tsung Chan$^1$, Pei-Ming Ho$^2$, Jen-Chi Lee$^3$,
Shunsuke Teraguchi$^4$, Yi Yang$^3$}
\vskip 15pt
{\small $^1$Physics Division,
National Center for Theoretical Sciences,
Hsinchu, Taiwan, R.O.C.}\\
{\small $^2$Department of Physics,
National Taiwan University,
Taipei, Taiwan, R.O.C.}\\
{\small $^3$Department of Electrophysics,
National Chiao-Tung University,
Hsinchu, Taiwan, R.O.C.}\\
{\small $^4$Physics Division,
National Center for Theoretical Sciences,
Taipei, Taiwan, R.O.C.}\\
\vskip .2in
\sffamily{
ctchan@phys.cts.nthu.edu.tw\\
pmho@ntu.edu.tw\\
jcclee@cc.nctu.edu.tw\\
teraguch@phys.ntu.edu.tw\\
yiyang@mail.nctu.edu.tw}
\vspace{60pt}
\end{center}
\begin{abstract}
We study the implication of decoupling zero-norm states
in the high-energy limit,
for the 26 dimensional bosonic open string theory.
Infinitely many linear relations among 4-point functions
are derived algebraically, and their unique solution is found.
Equivalent results are also obtained by taking the high-energy limit
of Virasoro constraints,
and as an independent check,
we compute all 4-point functions of 3 tachyons and an arbitrary massive state
by saddle-point approximation.
\end{abstract}
\end{titlepage}
\setcounter{footnote}{0}

\section{Introduction}

\subsection{Motivation}

The high-energy limit of string theory is an old subject
\cite{GM,Gross,GrossManes}. Its motivation is to find the hypothetical hidden
symmetry. It is believed that the higher-spin gauge fields in string theory
receive their masses via a spontaneous symmetry breaking mechanism
\cite{Gross,LEO}. It is then natural to expect that the gauge fields become
effectively massless and the symmetry is restored in the high-energy limit.
The symmetry might also explain mysterious correspondences such as dualities
and holography.

This subject has been attacked from various directions. One can take the
high-energy limit of the worldsheet theory and use it to define string theory
in the high-energy limit \cite{WS}. However there is a lot of ambiguity in
taking the high-energy limit of a worldsheet. It is not clear which choice of
the high-energy worldsheet action is the one suitable for the purpose of
studying symmetries.

Another approach is to study consistent higher spin gauge theories \cite{HS}.
It turns out that there is no consistent scale-free nontrivial interactions in
the flat spacetime. Usually people study these theories in the AdS background.
In general we need a characteristic length scale to define consistent
interactions. The string scale $\a^{\prime}$ can not be used because by
definition $\a^{\prime}\rightarrow\infty$ in the high-energy limit.

The third approach is to examine correlation functions in string theory and
look for meaningful patterns in the high-energy limit \cite{GM, Gross,
GrossManes}. This is the approach that we will take in this work. We consider
the bosonic open string theory and compare correlation functions which are
different from each other by a single vertex at the same mass level. We claim
that their relative ratios are completely determined at the leading order for
any mass level. This should be viewed as the strongest signal there ever was
for a symmetry in the high-energy limit.

The main idea of our approach is that the decoupling of zero-norm states
strongly restricts the dynamics of string theory \cite{GSW}. One can show
\cite{KaoLee,CLYang} that off-shell gauge
transformations of Witten's string field theory, after imposing the no-ghost
condition, are identical to the on-shell stringy gauge symmetries generated by
two types of zero-norm states.
In view of the dictating role of gauge symmetry in Witten's SFT,
we believe that
the existence of the huge gauge symmetry represented by zero-norm states
\cite{Lee,JCLee} fixes the theory uniquely.
In particular, we will show in this paper that by taking a self-consistent
high-energy limit \cite{ChanLee1,ChanLee2,CHL} of arbitrary four-point functions, it will allow us
to express them at the leading order in terms of those of tachyons.
We will also take the 2D string as an example to establish the connection
between the high-energy limit of zero-norm states and the hidden symmetry.
In 2D string, we will show that the high-energy limit of $w_\infty$ algebra generated by 2D zero-norm states
\cite{ChungLee} can be identified with the $w_\infty$ symmetry of 2D string \cite{Winfinity}.

\subsection{Review}

\label{Review}

We will focus on 4-point functions in this work, although our discussion can
be generalized to higher point correlation functions. Due to Poincare
symmetry, a 4-point function is a function of merely two parameters. Viewing a
4-point function as the scattering amplitude of a two-body scattering process,
one can choose the two parameters to be $E$ (one half of the center of mass
energy for the incoming particles i.e., particles 1 and 2 in Fig. 1, and
$\phi$ (the scattering angle between particles 1 and 3). For convenience we
will take the center of mass frame and put the momenta of particles 1 and 2
along the $X^{1}$-direction, with the momenta of particles 3 and 4 on the
$X^{1}-X^{2}$ plane. Readers can refer to the appendix \ref{appendix} for
expressions of the kinematic variables in this frame.

\begin{figure}[ptb]
\label{scattering} \setlength{\unitlength}{3pt}
\par
\begin{center}
\begin{picture}(100,100)(-50,-50)
{\large
\put(45,0){\vector(-1,0){42}} \put(-45,0){\vector(1,0){42}}
\put(2,2){\vector(1,1){30}} \put(-2,-2){\vector(-1,-1){30}}
\put(25,2){$k_1$} \put(-27,2){$k_2$} \put(11,20){$-k_3$}
\put(-24,-15){$-k_4$}
\put(40,0){\vector(0,-1){10}} \put(-40,0){\vector(0,1){10}}
\put(26,26){\vector(-1,1){7}} \put(-26,-26){\vector(1,-1){7}}
\put(36,-16){$e^{T}(1)$} \put(-44,15){$e^{T}(2)$}
\put(15,36){$e^{T}(3)$} \put(-18,-35){$e^{T}(4)$}
\qbezier(10,0)(10,4)(6,6) \put(12,4){$\phi$}
\put(-43,-45){Fig.1 Kinematic variables in the center of mass frame} }
\end{picture}
\end{center}
\end{figure}

The high-energy limit under consideration is
\begin{equation}
\alpha^{\prime}E^{2} \rightarrow\infty, \quad\phi= \mbox{fixed}.
\end{equation}
Based on the saddle-point approximation of Gross and Mende \cite{GM}, Gross
and Manes \cite{GrossManes} computed the high-energy limit of 4-point
functions in the bosonic open string theory. To explain their result, let us
first define our notations and conventions. For a particle of momentum $k$, we
define an orthonormal basis of polarizations $\{e^{P}, e^{L}, e^{T_{i}} \}$.
The momentum polarization $e^{P}$ is proportional to $k$, the longitudinal
polarization $e^{L}$ is the space-like unit vector whose spatial component is
proportional to that of $k$, and $e^{T_{i}}$ are the space-like unit-vectors
transverse to the spatial momentum. As an example, for $k$ pointing along the
$X^{1}$-direction,
\begin{equation}
k = (k^{0}, k^{1}, k^{2}, \cdots, k^{25}) = (E, p, 0, \cdots, 0), \hspace{1cm}
p > 0,
\end{equation}
the basis of polarization is
\begin{equation}
e^{P} = \frac{1}{m}(\sqrt{p^{2}+m^{2}}, p, 0, 0, \cdots, 0), \hspace{0.3cm}
e^{L} = \frac{1}{m}(p,\sqrt{p^{2}+m^{2}}, 0, 0, \cdots, 0), \hspace{0.3cm}
e^{T_{i}} = (0,0,\cdots,1, \cdots),
\end{equation}
where $m$ is the mass of the particle. In general, $e^{T_{i}}$ (for $i = 3,
\cdots, 25$) is just the unit vector in the $X^{i}$-direction, and the
definitions of $e^{P}, e^{L}$ and $e^{T_{2}}$ will depend on the motion of the
particle. For $e^{T_{2}}$, which is parallel to the scattering plane, we
denote it by $e^{T}$ (see Fig. 1 and the appendix \ref{appendix}). The
orientations of $e^{T_{2}}$ for each particle are fixed by the right-hand
rule, $\vec{k} \times e^{T_{2}} = e^{T_{3}}$, where $\vec{k}$ is the spatial
momentum of 4-vector $k$. We will use the notation $\del^{n} X^{A} \equiv
e^{A} \cdot\del^{n} X$ for $A = P, L, T, T_{i}$.

Each vertex is a polynomial of $\{\del^{n} X^{A}\}$ times the exponential
factor $\exp(ik\cdot X)$. Among all possible choices of polarizations for the
4 vertices in a 4-point function, we now argue that only the polarizations $L$
and $T$ need to be considered. The polarization $P$ can be gauged away using
zero-norm states \cite{CLYang}. To see why we can ignore all $T_{i}$'s except
$T$, we note that a prefactor $\del^{n} X^{A}$ can be contracted with the
exponent $i k \cdot X$ of another vertex. The contribution of this contraction
to a scattering amplitude is proportional to $k^{A} \sim E$. If $k^{A} \neq0$
(i.e., if $A = L$ or $T$), this is much more important in the high-energy
limit than a contraction with another prefactor $\del^{m} X^{B}$, which gives
$\eta^{AB} \sim E^{0}$. Therefore, if all polarizations in all vertices are
chosen to be either $L$ or $T$, the resulting 4-point function will dominate
over other choices of polarizations.

According to the zeroth-order saddle-point approximation \cite{GM}, the
leading order of a 4-point function is
\begin{equation}
\langle V(k_{1}) V(k_{2}) V(k_{3}) V(k_{4}) \rangle\sim\left(  \prod_{a=1}^{4}
v_{a}\right)  {\cT}.
\end{equation}
Here ${\cT}$ is the correlation function of 4 tachyons,
according to the following rules
\begin{align}
\frac{1}{(n-1)!} \del^{n} X^{T}  &  \rightarrow i \frac{E\sin\phi}%
{(-\lam)^{n}},\label{XT}\\
\frac{1}{(n-1)!} \del^{n} X^{L}  &  \rightarrow-i \frac{\lam^{n-1} - 1}{\lam -
1}\frac{E^{2} \sin^{2}\phi}{2M (-\lam)^{n}}, \label{XL}%
\end{align}
where\footnote
{However, see correction in Reference \cite{CHL}.}
\begin{equation}
\lam = \sin^{2}\frac{\phi}{2}.
\end{equation}
A peculiar feature of (\ref{XL}) is that it vanishes for $n = 1$
\begin{equation}
\del X^{L} \rightarrow0. \label{XL1}%
\end{equation}
Does this mean that all 4-point functions involving $\del X^{L}$ are
subleading? The answer depends on how we define the notion of ``subleading''.
What is the reference 4-point function to be compared with?

Note that it is impossible to have a universal notion of ``leading order'' for
all 4-point functions, since 4-point functions with vertices at higher mass
levels typically dominate over those with vertices at lower mass levels in the
high-energy limit. Having a well defined limit for all correlation functions
is also contradictory to the fact that there is no consistent interacting
higher-spin gauge field theory in the flat spacetime.

We propose that the appropriate question to ask is: How will a 4-point
function change if we replace a vertex by another vertex at the same mass
level? For example, we can replace a factor $\del X^{L}$ by $\del X^{T}$ in
one of the vertices. Naively, the former with $\del X^{L}$ should dominate
over the latter in the high-energy limit since the components of $e^{L}$ scale
as $E^{1}$ and those of $e^{T}$ scale as $E^{0}$. However, Eq. (\ref{XL1})
tells us that $\del X^{L}$ does not live up to this expectation. Nevertheless,
it does not imply that $\del X^{L}$ is subleading compared with $\del X^{T}$.
It depends on the rest of the terms in the prefactor.

The purpose of our work is to find linear relations among 4-point functions in
the high-energy limit as a signature of the hidden symmetry proposed in
\cite{GM}. This is not manifest in the saddle-point result (\ref{XT}),
(\ref{XL}), which holds for the \emph{naive} leading order (0-th order
saddle-point approximation). One of the main reasons is that they ignored
vertices involving $\del X^{L}$. Another main reason is that they did not try
to determine the precise power factor of $E$ \cite{CHL} in a 4-point function
(a higher order correction to the saddle-point approximation is needed), which
is crucial for examining linear relations among 4-point functions.

The proper notion of ``leading order'' in the high-energy limit such that
linear relations among 4-point functions can be established was first
discovered in \cite{ChanLee1} and \cite{ChanLee2}. A purely algebraic approach
utilizing zero-norm states was developed there to derive the linear relations
explicitly for the first few mass levels. This approach was further explored
to obtain stronger results, and its connection/difference from the
saddle-point computation was explained in detail in \cite{CHL}.

The central idea behind the algebraic approach used in \cite{ChanLee1},
\cite{ChanLee2} and \cite{CHL} was the decoupling of zero-norm states (i.e.,
the requirement of gauge invariance). A crucial step in the derivation is to
replace the polarization $e^{P}$ by $e^{L}$ in the zero-norm states. It is
assumed that, while zero-norm states decouple at all energies, the replacement
leads to states that are decoupled at high energies.

The new achievements of this paper include:

\begin{enumerate}
\item We studied the implication of the decoupling of zero-norm states in the
high-energy limit. The 4-point functions at the leading order are specified
for all mass levels.

\item Infinite linear relations among 4-point functions are obtained for all
mass levels. The ratio of any two 4-point functions at the leading order is
uniquely determined.

\item We developed a ``dual'' description of the algebraic approach using the
spurious states, which is to take high-energy limit of the Virasoro constraints.

\item Using saddle-point approximation, we explicitly computed all 4-point
functions for 3 tachyons and one arbitrary massive state, and verified the
linear relations obtained via algebraic methods.
\end{enumerate}

\section{Main results}

For brevity, we will refer to all 4-point functions different from each other
by a single vertex at the same mass level as a ``\emph{family}''. When we
compare members of a family, we only need to specify the vertex which is changed.

A 4-point function will be said to be \emph{at the leading order} if it is not
subleading to any of its \emph{siblings}. We will ignore those that are not at
the leading order. Our aim is to find the numerical ratios of all 4-point
functions in the same family at the leading order. Apparently, there are more
4-point functions at the leading order at higher mass levels. Our goal may
seem insurmountable at first sight.

Saving the derivation for later, we give our main results here. A 4-point
function is at the leading order if and only if the vertex $V$ under
comparison is a linear combination of vertices of the form
\begin{equation}
V^{(n,m,q)}(k) = \left(  \del X^{T}\right)  ^{n-m-2q} \left(  \del X^{L}%
\right)  ^{m} \left(  \del^{2} X^{L}\right)  ^{q} e^{ik\cdot X} ,
\end{equation}
where
\begin{equation}
\quad n \geq m+2q, \quad m, q \geq0.
\end{equation}
The corresponding states are of the form
\begin{equation}
\left(  \a^{T}_{-1}\right)  ^{n-m-2q} \left(  \a^{L}_{-1}\right)  ^{m} \left(
\a^{L}_{-2}\right)  ^{q} |0,k\rangle.
\end{equation}
The mass squared is $2(n-1)$. All other states involving $\a^{T}_{-2},
\a^{A}_{-3}, \cdots$ are subleading.

Using the notation\footnote{More rigorously, $V_{2}$ needs to be a physical
state in order for the correlation function to be well-defined. We should keep
in mind that our results should be applied to suitable linear combinations of
(9), possibly together with subleading states, to satisfy Virasoro
constraints.}
\begin{equation}
\cT^{(n,m,q)} = \langle V_{1} V^{(n,m,q)}(k) V_{3} V_{4} \rangle,
\end{equation}
all linear relations among different choices of $V^{(m,n,q)}$ (obtained from
the decoupling of spurious states at high energies) can be solved by the
simple expression
\begin{align}
\lim_{E \rightarrow\infty} \frac{\cT^{(n,2m,q)}}{\cT^{(n,0,0)}}  &  = \left(
-\frac{1}{\hat{m}}\right)  ^{2m+q} \left(  \frac{1}{2}\right)  ^{m+q}
(2m-1)!!,\label{main}\\
\lim_{E \rightarrow\infty} \frac{\cT^{(n,2m+1,q)}} {\cT^{(n,0,0)}}  &  = 0,
\end{align}
where $\hat{m}=\sqrt{2(n-1)}$. This formula tells us how to trade $\del
X^{L}$ and $\del^{2} X^{L}$ for $\del X^{T}$, so that all 4-point functions
can be related to the one involving only $\del X^{T}$ in $V_{2}$. The formula
above applies equally well to all vertices.

Since we know the value of a representative 4-point function \cite{ChanLee1,
CHL}
\begin{equation}
\mathcal{T}_{n_{1}n_{2}n_{3}n_{4}}^{T^{1}\cdot\cdot T^{2}\cdot\cdot T^{3}
\cdot\cdot T^{4}\cdot\cdot}= (-1)^{n_{2} + n_{4}} [ 2 E^{3}\sin\phi
_{CM}]^{\Sigma n_{i}} \mathcal{T}(\Sigma n_{i}), \label{TN}%
\end{equation}
where
\begin{align}
\mathcal{T}(n)  &  =\sqrt{\pi}(-1)^{n-1}2^{-n}E^{-1-2n}(\sin\frac{\phi_{CM}%
}{2})^{-3}(\cos\frac{\phi_{CM}}{2})^{5-2n}\nonumber\\
&  \times\exp(-\frac{s\ln s+t\ln t-(s+t)\ln(s+t)}{2})
\end{align}
is the high-energy limit of $\frac{\Gamma(-\frac{s}{2}-1)\Gamma(-\frac{t}%
{2}-1)}{\Gamma(\frac{u}{2}+2)}$ with $s+t+u=2 \Sigma n_{i} -8$, and we have
calculated it up to the next leading order in $E$. In Eq.(\ref{TN}), $n_{i}$
is the number of $T^{i}$ of the $i-th$ vertex operators and $T^{i}$ is the
transverse direction of the $i-th$ particle. We can immediately write down the
explicit expression of a 4-point function if all vertices are nontrivial at
the leading order.

\section{Linear relations among 4-point functions}

\label{Spurious}

Before we go on, we recall some terminology used in the old covariant
quantization. A state $|\psi\rangle$ in the Hilbert space is \emph{physical}
if it satisfies the Virasoro constraints
\begin{equation}
\left(  L_{n}-\delta_{n}^{0}\right)  |\psi\rangle=0,\quad n\geq0.
\end{equation}
Since $L_{n}^{\dagger}=L_{-n}$, states of the form
\begin{equation}
L_{-n}|\chi\rangle
\end{equation}
are orthogonal to all physical states, and they are called \emph{spurious
states}. \emph{Zero-norm states} are spurious states that are also physical.
They correspond to gauge symmetries. In the old covariant first quantization
spectrum of open bosonic string theory, the solutions of physical state
conditions include positive-norm propagating states and two types of zero-norm
states. The latter are \cite{GSW}
\begin{align}
\text{Type I}:L_{-1}\left\vert x\right\rangle ,  &  \text{ where }%
L_{1}\left\vert x\right\rangle =L_{2}\left\vert x\right\rangle =0,\text{
}L_{0}\left\vert x\right\rangle =0;\\
\text{Type II}:(L_{-2}+\frac{3}{2}L_{-1}^{2})\left\vert \widetilde
{x}\right\rangle ,  &  \text{ where }L_{1}\left\vert \widetilde{x}%
\right\rangle =L_{2}\left\vert \widetilde{x}\right\rangle =0,\text{ }%
(L_{0}+1)\left\vert \widetilde{x}\right\rangle =0.
\end{align}

In this section we derive the linear relations among all amplitudes in the
same family by taking the high-energy limit of zero-norm states (HZNS).
Solutions of HZNS for some low lying mass level are presented in the appendix
B. The first step in the derivation is to identify the class of states that
are relevant, i.e., those at the leading order. As we explained in sec.
\ref{Review}, we only need to consider the polarizations $e^{T}$ and $e^{L}$.

To get a rough idea about how each vertex operator scales with $E$ in the
high-energy limit, we associate a naive dimension to each prefactor $\del^{m}
X^{A}$ according to the following rule
\begin{equation}
\del^{m} X^{T} \rightarrow1, \quad\del^{m} X^{L} \rightarrow2.
\end{equation}
The reason is the following. Each factor of $\del^{m} X^{\mu}$ has the
possibility of contracting with the exponent $ik_{i} \cdot X$ of another
vertex operator so that it scales like $E$ in the high-energy limit.
Furthermore, components of the polarization vectors $e^{T}$ and $e^{L}$ scale
with $E$ like $E^{0}$ and $E^{1}$, respectively.

When we compare vertex operators at the same mass level, the sum of all the
integers $m$ in $\del^{m} X^{A}$ is fixed. Roughly speaking, it is
advantageous to have many $\del X^{A}$ than having fewer number of $\del^{m}
X^{A}$ with $m > 1$. For example, at the first massive level, the vertex
operator $\del X^{T} \del X^{T} e^{ik\cdot X}$ has a larger naive dimension
than $\del^{2} X^{T} e^{ik\cdot X}$.

The counting of the naive dimension does not take into consideration the
possibility that the coefficient of the leading order term happens to vanish
by cancellation. The true dimension of a vertex operator can be lower than its
naive dimension, although the reverse never happens.

Through experiences accumulated from explicit computations \cite{ChanLee1,
ChanLee2, CHL}, we find that the highest spin vertex
\begin{equation}
(\del X^{T})^{n} e^{ik\cdot X} \leftrightarrow\left(  \a_{-1}^{T}\right)  ^{n}
|0,k\rangle
\end{equation}
is always at the leading order in its family. Since the naive dimension of
this state equals its true dimension, any state with a lower naive dimension
than this vertex operator can be ignored. This implies that we can immediately
throw away a lot of vertex operators at each mass level, but there are still
many left. The problem is that, although there are disadvantages to have
$\del^{m} X^{T}$ with $m \geq2$ or $\del^{m} X^{L}$ with $m \geq3$ compared
with having $\left(  \del
X^{T}\right)  ^{m}$, it may be possible that having extra factors of
$\del X^{L}$, which has a higher naive dimension than $\del X^{T}$, can
compensate the disadvantage of these factors. However, explicit computations
at the first few massive levels showed that this never happens.

We will now argue why this is generically true, and show in this subsection
that the only states that will survive the high-energy limit at level $n$ are
of the form
\begin{equation}
\label{relevant}|n, 2m, q\rangle\equiv\left(  \a^{T}_{-1}\right)
^{n-2m-2q}\left(  \a^{L}_{-1}\right)  ^{2m}\left(  \a_{-2}^{L}\right)  ^{q}
|0; k\rangle.
\end{equation}

Our argument is essentially based on the decoupling of zero-norm states in the
high-energy limit. However, note that when we take the high-energy limit, that
is, when we replace $e^{P}$ by $e^{L}$  and
ignore subleading terms in the $1/E^2$ expansion\footnote
{Strictly speaking, we need to
justify the replacement $e^{P} \rightarrow e^{L}$. This is not totally
trivial, and will be treated with more rigor in a forthcoming paper.},
the zero-norm states become
positive norm states, although we will still call them ``high-energy zero-norm
states'', or HZNS for short. (This is why it is possible to derive relations
among positive-norm physical states by taking the high-energy limit of Ward
identities. See Appendix B for examples.) As such, it is not essential to
maintain the zero-norm condition, and we can simply take the high-energy limit
of spurious states. It can be shown that, as far as the final results are
concerned, the decoupling of those spurious states we are going to use in this
paper are equivalent to the decoupling of high-energy zero-norm states (HZNS).

Thanks to the Virasoro algebra, we only need two Virasoro operators
\begin{align}
L_{-1}  &  = \frac{1}{2} \sum_{n\in\Z} \a_{-1+n}\cdot\a_{-n} = \hat
m\ap_{-1}^{P}+\ap_{-2}\cdot\ap_{1}+\cdots,\\
L_{-2}  &  = \frac{1}{2} \sum_{n\in\Z} \a_{-2+n}\cdot\a_{-n} = \frac{1}{2}
\ap_{-1}\cdot\ap_{-1}+\hat m\ap_{-2}^{P}+\ap_{-3}\cdot\ap_{1}+\cdots
\end{align}
to generate all spurious states. Here $\hat m$ is the mass operator, i.e.,
$\hat{m}^{2} = - k^{2}$ when acting on the state $|0, k\rangle$.

\subsection{Irrelevance of other states}

To prove that only states of the form (\ref{relevant}) are at the leading
order, we shall prove that $(i)$ any state which has an odd number of
$\a_{-1}^{L}$ is irrelevant (i.e., subleading in the high-energy limit), and
$(ii)$ any state involving a creation operator whose naive dimension is less
than its mode index $n$, i.e., states belonging to
\begin{equation}
\{\a_{-n}^{L},\;\;n>2;\quad\a_{-m}^{T},\;\;m>1\} \label{irrset}%
\end{equation}
is also irrelevant. We proceed by mathematical induction.

First we prove that any state which has a single factor of $\a_{-1}^{L}$ is
irrelevant, and that any state with two $\a_{-1}^{L}$'s is irrelevant if it
contains an operator of naive dimension less than its index.

Consider the HZNS $L_{-1} \chi$ where $\chi$ is any state without any
$\a_{-1}^{L}$, and it is at level $(n-1)$. Note that, except $\a_{-1}^{L}$,
the naive dimension of an operator is always less than or equal to its index
(we exclude $\a_{-1}^{P}$ as mentioned above). This means that the naive
dimension of $\chi$ is less than or equal to $(n-1)$. Since we know that at
level $n$, the state Eq.(\ref{relevant}) has true dimension $n$, when
computing $L_{-1} \chi$ in the high-energy limit, we can ignore everything
with naive dimension less than $n$. This means that we need $L_{-1}$ to
increase the naive dimension of $\chi$ by no less than 1. In the high-energy
limit of $L_{-1}$
\begin{equation}
\label{HL-1}L_{-1} \rightarrow\hat m\a_{-1}^{L} + \a_{-2}^{L} a_{1}^{L} +
\a_{-2}^{T} \a_{1}^{T} + \cdots,
\end{equation}
only the first term will increase the naive dimension of $\chi$ by 1. All the
rest do not change the naive dimension. This means that, to the leading
order,
\begin{equation}
L_{-1} \chi\sim\hat{m} \a_{-1}^{L} \chi.
\end{equation}
This is a state with a single factor of $\a_{-1}^{L}$ and it is a HZNS, so it
should be decoupled in the high-energy limit.

Now consider an arbitrary state $\chi$ at level $(n-1)$ which has a single
factor of $\a^{L}_{-1}$. If $\chi$ involves any operator whose naive dimension
is less than its index, the naive dimension of $\chi$ is at most $(n-1)$. In
the high-energy limit
\begin{equation}
\label{L-1chi}L_{-1} \chi\rightarrow\hat{m} \a_{-1}^{L} \chi+ \a_{-2}^{L}
\a_{1}^{L} \chi+ \cdots,
\end{equation}
except the first two terms, all other terms are irrelevant because they
contain a single factor of $\a_{-1}^{L}$. As the second term has a naive
dimension $(n-1)$ and can be ignored, we conclude that $\a_{-1}^{L} \chi$ is irrelevant.

The next step in mathematical induction is to show that if (a) states with
$(2m-1)$ factors of $\a^{L}_{-1}$ are irrelevant, and (b) states with $2m$
factors of $\a^{L}_{-1}$ are still irrelevant if it also contains any of the
operators in (\ref{irrset}), then we can prove that both statements are also
valid for $m\rightarrow m+1$.

Suppose $\chi$ is an arbitrary state at level $(n-1)$ which has $2m$ factors
of $\a^{L}_{-1}$'s. The high-energy limit of $L_{-1}\chi$ is given by
(\ref{L-1chi}). The second term has $(2m-1)$ factors of $\a^{L}_{-1}$ and is
irrelevant. The rest of the terms, except the first, are irrelevant because
they contains at least one operator from the set (\ref{irrset}). Hence the
first term is a HZNS and is irrelevant. We have proved our first claim for
$(m+1)$, i.e., a state with $(2m+1)$ factors of $\a^{L}_{-1}$ decouple at high energies.

Similarly, consider the case when $\chi$ is at level $(n-1)$ and has $(2m-1)$
factors of $\a_{-1}^{L}$. Furthermore we assume that it involves operators
from the set (\ref{irrset}). Then the first term in (\ref{L-1chi}) is what we
want to prove to be irrelevant. The second term is irrelevant because we have
just proved that a state with $(2m+1)$ factors of $\a^{L}_{-1}$ is irrelevant.
The rest of the terms are irrelevant because they have $(2m-1)$ $\a^{L}_{-1}%
$'s. Thus we conclude that both claims are correct for $m+1$ as well. The
mathematical induction is complete.

\subsection{Linear relations}

According to the previous subsection, only states of the form (\ref{relevant})
are relevant in the high-energy limit. The mass of the state is $\sqrt
{2(n-1)}$. The 4-point function associated with $|n, m, q\rangle$ will be
denoted $\mathcal{T}^{(n,m,q)}$. The aim of this subsection is to find the
ratio between a generic $\mathcal{T}^{(n,m,q)}$ and the reference 4-point
function, which is taken to be $\mathcal{T}^{(n,0,0)}$.

Consider the HZNS
\begin{equation}
L_{-1}|n-1, 2m-1, q\rangle\simeq\hat{m} |n, 2m, q\rangle+ (2m-1) |n, 2m-2, q+1
\rangle,
\end{equation}
where many terms are omitted because they are not of the form (\ref{relevant}%
). This implies that
\begin{equation}
\mathcal{T}^{(n, 2m, q)} = - \frac{2m-1}{\hat{m}} \mathcal{T}^{(n, 2m-2,
q+1)}.
\end{equation}
Using this relation repeatedly, we get
\begin{equation}
\label{1}\mathcal{T}^{(n, 2m, q)} = \frac{(2m-1)!!}{(- \hat{m})^{m}}
\mathcal{T}^{(n, 0, m+q)},
\end{equation}
where the double factorial is defined by $(2m-1)!! = \frac{(2m)!}{2^{m} m!}$.

Next, consider another class of HZNS
\begin{equation}
L_{-2}|n-2, 0, q\rangle\simeq\frac{1}{2} |n, 0, q\rangle+ \hat{m} |n, 0,
q+1\rangle.
\end{equation}
Again, irrelevant terms are omitted here. From this we deduce that
\begin{equation}
\mathcal{T}^{(n, 0, q+1)} = -\frac{1}{2 \hat{m}} \mathcal{T}^{(n, 0, q)},
\end{equation}
which leads to
\begin{equation}
\label{2}\mathcal{T}^{(n, 0, q)} = \frac{1}{(-2 \hat{m})^{q}} \mathcal{T}^{(n,
0, 0)}.
\end{equation}

Our main result (\ref{main}) is an immediate result of combining (\ref{1}) and
(\ref{2}).

\section{Linear relations from Virasoro constraints}

In this section we will establish a \textquotedblleft dual
description\textquotedblright\ of our approach explained above. The notion
dual to the decoupling of high-energy zero-norm states is Virasoro constraints.

Let us briefly explain how to proceed. First write down a state at a given
mass level as linear combination of states of the form Eq.(11) with
undetermined coefficients, which are interpreted as the Fourier components of
spacetime fields. Requiring that the Virasoro generators $L_{1}$ and $L_{2}$
annihilate the state implies several linear relations on the coefficients. The
linear relations can then be solved to obtain ratios among all fields.

To compare the results of the two dual descriptions, we note that the
correlation functions can be interpreted as source terms for the particle
corresponding to a chosen vertex. Thus the ratios among sources should be the
same as the ratios among the fields, since all fields of the same mass have
the same propagator. However, some care is needed for the normalization of the
field variables. One should use BPZ conjugates to determine the norm of a
state and normalize the fields accordingly.

\subsection{Examples}

To illustrate how Virasoro constraints can be used to derive linear relations
among scattering amplitudes at high energies, we give some explicit examples
in this subsection. We will calculate the proportionality constants among high
energy scattering amplitudes of different string states up to mass levels
$m^{2}=8$. The results are of course consistent with those of previous work
\cite{ChanLee1,ChanLee2} using high-energy zero-norm states.

\subsubsection{$m^{2}=4$}

The most general form of physical states at mass level $m^{2}=4$ are given by
\begin{equation}
\lbrack\epsilon_{\mu\nu\lambda}\alpha_{-1}^{\mu}\alpha_{-1}^{\nu}\alpha
_{-1}^{\lambda}+\epsilon_{(\mu\nu)}\alpha_{-1}^{\mu}\alpha_{-2}^{\nu}%
+\epsilon_{\lbrack\mu\nu]}\alpha_{-1}^{\mu}\alpha_{-2}^{\nu}+\epsilon_{\mu
}\alpha_{-3}^{\mu}]|0,k\rangle. \label{49}%
\end{equation}
The Virasoro constraints are
\begin{align}
\epsilon_{(\mu\nu)}+\frac{3}{2}k^{\lambda}\epsilon_{\mu\nu\lambda}  &
=0,\label{50}\\
-k^{\nu}\epsilon_{\lbrack\mu\nu]}+3\epsilon_{\mu}-\frac{3}{2}k^{\nu}%
k^{\lambda}\epsilon_{\mu\nu\lambda}  &  =0,\label{51}\\
2k^{\nu}\epsilon_{\lbrack\mu\nu]}+3\epsilon_{\mu}-3(k^{\nu}k^{\lambda}%
-\eta^{\nu\lambda})\epsilon_{\mu\nu\lambda}  &  =0. \label{52}%
\end{align}
By replacing $P$ by $L$, and ignoring irrelevant states (we have justified
this in sec. \ref{Spurious} for the high-energy limit), one easily gets
\begin{equation}
\epsilon_{TTT}:\epsilon_{(LLT)}:\epsilon_{(LT)}:\epsilon_{\lbrack
LT]}=8:1:3:-3. \label{53}%
\end{equation}
After including the normalization factor of the field variables \footnote{The
normalization factors are determined by the inner product of a state with its
BPZ conjugate.} and the appropriate symmetry factors, one ends up with%
\begin{align}
&  \mathcal{T}_{TTT}:\mathcal{T}_{(LLT)}:\mathcal{T}_{(LT)}:\mathcal{T}%
_{[LT]}\nonumber\\
&  =6\epsilon_{TTT}:6\epsilon_{(LLT)}:-2\epsilon_{(LT)}:-2\epsilon_{\lbrack
LT]}=8:1:-1:1. \label{54}%
\end{align}
Here the definitions of $\mathcal{T}_{TTT}, \mathcal{T}_{(LLT)},
\mathcal{T}_{(LT)}, \mathcal{T}_{[LT]}$ and similar amplitudes hereafter can
be found in \cite{ChanLee1,ChanLee2} and the result obtained is consistent
with the previous zero-norm state calculation in \cite{ChanLee1} or
Eq.(\ref{main}).

\subsubsection{$m^{2}=6$}

The most general form of physical states at mass level $m^{2}=6$ are given by
\begin{align}
&  [\epsilon_{\mu\nu\lambda\sigma}\alpha_{-1}^{\mu}\alpha_{-1}^{\nu}%
\alpha_{-1}^{\lambda}\alpha_{-1}^{\sigma}+\epsilon_{(\mu\nu\lambda)}%
\alpha_{-1}^{\mu}\alpha_{-1}^{\nu}\alpha_{-2}^{\lambda}+\epsilon_{\mu
\nu,\lambda}\alpha_{-1}^{\mu}\alpha_{-1}^{\nu}\alpha_{-2}^{\lambda}\nonumber\\
&  +\epsilon_{(\mu\nu)}^{(1)}\alpha_{-1}^{\mu}\alpha_{-3}^{\nu}+\epsilon
_{\lbrack\mu\nu]}^{(1)}\alpha_{-1}^{\mu}\alpha_{-3}^{\nu}+\epsilon_{(\mu\nu
)}^{(2)}\alpha_{-2}^{\mu}\alpha_{-2}^{\nu}+\epsilon_{\mu}\alpha_{-4}^{\mu
}]|0,k\rangle, \label{55}%
\end{align}
where $\epsilon_{\mu\nu,\lambda}$ represents the mixed symmetric spin three
states, that is, one first symmetrizes $\mu\nu$ and then anti-symmetrizes
$\mu\lambda.$ The Virasoro constraints are calculated to be
\begin{align}
2k^{\sigma}\epsilon_{(\mu\nu\lambda\sigma)}+\epsilon_{(\mu\nu\lambda)}  &
=0,\label{56}\\
2k^{\lambda}\epsilon_{(\mu\nu\lambda)}+k^{\lambda}(\epsilon_{\lambda\mu,\nu
}+\epsilon_{\mu\lambda,\nu})+3(\epsilon_{(\mu\nu)}^{(1)}+\epsilon_{\lbrack
\mu\nu]}^{(1)})+4\epsilon_{(\mu\nu)}^{(2)}  &  =0,\label{57}\\
k^{\mu}\epsilon_{(\mu\nu)}^{(1)}+k^{\mu}\epsilon_{\lbrack\mu\nu]}%
^{(1)}+4\epsilon_{\nu}  &  =0,\label{58}\\
6\eta^{\lambda\sigma}\epsilon_{(\mu\nu\lambda\sigma)}+2k^{\lambda}%
\epsilon_{(\mu\nu\lambda)}+\frac{1}{2}k^{\lambda}(\epsilon_{\mu\nu,\lambda
}+\epsilon_{\nu\mu,\lambda})+3\epsilon_{(\mu\nu)}^{(1)}  &  =0,\label{59}\\
\eta^{\mu\nu}\epsilon_{(\mu\nu\lambda)}+\eta^{\mu\nu}\epsilon_{(\mu\nu
,\lambda)}+4k^{\mu}\varepsilon_{(\mu\nu)}^{(2)}+4\epsilon_{\lambda}  &  =0.
\label{60}%
\end{align}
In the high-energy limit, similar calculation as above gives
\begin{align}
&  \mathcal{T}_{(TTTT)}:\mathcal{T}_{(TTLL)}:\mathcal{T}_{(LLLL)}%
:\mathcal{T}_{TT,L}:\mathcal{T}_{(TTL)}:\mathcal{T}_{(LLL)}:\mathcal{T}%
_{(LL)}\nn\\
&  =4!\mathcal{\epsilon}_{(TTTT)}:4!\mathcal{\epsilon}_{(TTLL)}%
:4!\mathcal{\epsilon}_{(LLLL)}:-4\mathcal{\epsilon}_{TT,L}:-4\mathcal{\epsilon
}_{(TTL)} :-4\mathcal{\epsilon}_{(LLL)}:8\mathcal{\epsilon}_{(LL)}^{(2)} \nn\\
&  =16:\frac{4}{3}:\frac{1}{3}:-\frac{2\sqrt{6}}{3}:-\frac{4\sqrt{6}}{9}
:-\frac{\sqrt{6}}{9}:\frac{2}{3}, \label{61}%
\end{align}
which is consistent with the previous zero-norm state calculation in
\cite{ChanLee2} or Eq.(\ref{main}).

\subsubsection{$m^{2} = 8$}

The most general form of physical states at mass level $m^{2}=8$ are given by
(for simplicity, we neglect terms containing $\alpha_{-n}^{\mu}$ with
$n\geq3$)
\begin{align}
&  [\epsilon_{\mu\nu\lambda\sigma\rho}\alpha_{-1}^{\mu}\alpha_{-1}^{\nu}%
\alpha_{-1}^{\lambda}\alpha_{-1}^{\sigma}\alpha_{-1}^{\rho}+\epsilon_{(\mu
\nu\lambda\sigma)}\alpha_{-1}^{\mu}\alpha_{-1}^{\nu}\alpha_{-1}^{\lambda
}\alpha_{-2}^{\rho}+\epsilon_{(\mu\nu\lambda)}\alpha_{-1}^{\mu}\alpha
_{-2}^{\nu}\alpha_{-2}^{\lambda}\nonumber\\
&  +\epsilon_{\mu\nu\lambda,\sigma}\alpha_{-1}^{\mu}\alpha_{-1}^{\nu}%
\alpha_{-1}^{\lambda}\alpha_{-2}^{\rho}+\epsilon_{\mu,\nu\lambda}\alpha
_{-1}^{\mu}\alpha_{-2}^{\nu}\alpha_{-2}^{\lambda}]|0,k\rangle, \label{62}%
\end{align}
where $\epsilon_{\mu\nu\lambda,\sigma}$ represents the mixed symmetric spin
four states, that is, first symmetrizes $\mu\nu\lambda$ and then
anti-symmetrizes $\mu\sigma.$ Similar definition for the mixed symmetric spin
three states $\epsilon_{\mu,\nu\lambda}$. The Virasoro constraints are
calculated to be
\begin{align}
5k^{\sigma}\epsilon_{(\mu\nu\lambda\sigma\rho)}+2\epsilon_{(\mu\nu
\lambda\sigma)}  &  =0,\label{63}\\
3k^{\lambda}\epsilon_{(\mu\nu\lambda\sigma)}+\frac{1}{2}k^{\lambda}%
[(\epsilon_{\mu\nu\lambda,\sigma}+\epsilon_{\lambda\mu\nu,\sigma}%
+\epsilon_{\mu\lambda\nu,\sigma})+(\mu &  \leftrightarrow\nu)]\nonumber\\
+4\epsilon_{(\mu\nu\sigma)}+\epsilon_{\mu,\nu\sigma}+\epsilon_{\nu,\mu\sigma}
&  =0,\label{64}\\
k^{\mu}\epsilon_{(\mu\nu\lambda)}+\frac{1}{2}k^{\mu}(\epsilon_{\mu,\nu\lambda
}+\epsilon_{\mu,\lambda\nu})  &  =0,\label{65}\\
5\eta^{\rho\sigma}\epsilon_{(\mu\nu\lambda\sigma\rho)}+k^{\sigma}%
\epsilon_{(\mu\nu\lambda\sigma)}+\frac{1}{3}k^{\sigma}(\epsilon_{\mu\nu
\lambda,\sigma}+\epsilon_{\nu\lambda\mu,\sigma}+\epsilon_{\lambda\mu\nu
,\sigma})  &  =0,\label{66}\\
3\eta^{\nu\lambda}\epsilon_{(\mu\nu\lambda\sigma)}+\eta^{\nu\lambda}%
(\epsilon_{\mu\nu\lambda,\sigma}+\epsilon_{\lambda\mu\nu,\sigma}+\epsilon
_{\nu\lambda\mu,\sigma})+4k^{\lambda}\varepsilon_{(\mu\sigma\lambda
)}+2k^{\lambda}(\varepsilon_{\mu,\sigma\lambda}+\varepsilon_{\mu,\lambda
\sigma})  &  =0. \label{67}%
\end{align}
In the high-energy limit, similar calculation as above gives
\begin{align}
&  \mathcal{T}_{(TTTTT)}:\mathcal{T}_{(TTTL)}:\mathcal{T}_{(TTTLL)}%
:\mathcal{T}_{(TLLL)}:\mathcal{T}_{(TLLLL)}:\mathcal{T}_{(TLL)} :\mathcal{T}%
_{T,LL}:\mathcal{T}_{TLL,L}:\mathcal{T}_{TTT,L} \nn\\
&  =5!\mathcal{\epsilon}_{(TTTTT)}:3!\times2\mathcal{\epsilon}_{(TTTL)}%
:5!\mathcal{\epsilon}_{(TTTLL)}:3!\times2\mathcal{\epsilon}_{(TLLL)}%
:5!\mathcal{\epsilon}_{(TLLLL)}\nonumber\\
&  :8\mathcal{\epsilon}_{(TLL)}:8\mathcal{\epsilon}_{T,LL}:3!\times
2\mathcal{\epsilon}_{TLL,L}:3!\times2\mathcal{\epsilon}_{TTT,L}\nonumber\\
&  =32:\sqrt{2}:2:\frac{3\sqrt{2}}{16}:\frac{3}{8}:\frac{1}{3}:\frac{2}{3}
:\frac{\sqrt{2}}{16}:3\sqrt{2}, \label{68}%
\end{align}
which can be checked to be remarkably consistent with the results of
Eq.(\ref{main}) after Young tableaux decomposition.

\subsection{General mass levels}

In this section we calculate the ratios of string scattering amplitudes in the
high-energy limit for general mass levels by imposing Virasoro constraints.
The final result will, of course, be exactly the same as what we obtained by
requiring the decoupling of high-energy zero-norm states. In the presentation
here we use the notation of Young's tableaux.

We consider the general mass level $m^{2}=2\left(  n-1\right)  $. The most
general state can be written as%
\begin{equation}
\left\vert n\right\rangle =\left\{  \sum_{m_{j}}\overset{k}{\underset
{j=1}{\otimes}}\frac{1}{j^{m_{j}}m_{j}!}%
\begin{tabular}
[c]{|c|c|c|}\hline
$\mu_{1}^{j}$ & $\cdots$ & $\mu_{m_{j}}^{j}$\\\hline
\end{tabular}
\alpha_{-j}^{\mu_{1}^{j}\cdots\mu_{m_{j}}^{j}}\right\}  \left\vert
0,k\right\rangle , \label{general state}%
\end{equation}
where we defined the abbreviation
\begin{equation}
\alpha_{-j}^{\mu_{1}^{j}\cdots\mu_{m_{j}}^{j}}\equiv\alpha_{-j}^{\mu_{1}^{j}%
}\cdots\alpha_{-j}^{\mu_{m_{j}}^{j}},
\end{equation}
with $m_{j}$ is the number of the operator $\alpha_{-j}^{\mu}$. The summation
runs over all possible combinations of $m_{j}$'s with the constraints
\begin{equation}
\sum_{j=1}^{k}jm_{j}=n \quad\text{ and }\quad0\leq m_{j}\leq n,
\end{equation}
so that the total mass is $n$. It is obvious that $k$ is less or equal to $n$.
Since the upper indices $\left\{  \mu_{1}^{j}\cdots\mu_{m_{j}}^{j}\right\}  $
in $\alpha_{-j}^{\mu_{1}^{j}}\cdots\alpha_{-j}^{\mu_{m_{j}}^{j}}$ are
symmetric, we used the Young tableaux notation to denote the coefficients in
Eq.(\ref{general state}). The direct product $\otimes$ acts on the Young 
tableaux
in the standard way, for example
\begin{equation}%
\begin{tabular}
[c]{|c|c|}\hline
$1$ & $2$\\\hline
\end{tabular}
\otimes%
\begin{tabular}
[c]{|c|}\hline
$3$\\\hline
\end{tabular}
=
\begin{tabular}
[c]{|c|c|c|}\hline
$1$ & $2$ & $3$\\\hline
\end{tabular}
\oplus%
\begin{tabular}
[c]{|c|c}\hline
$1$ & \multicolumn{1}{|c|}{$2$}\\\hline
$3$ & \\\cline{1-1}%
\end{tabular}
.
\end{equation}
Finally, $1/\left(  j^{m_{j}}m_{j}!\right)  $ are the normalization factors.
To be clear, for example $n=4$, the state can be written as
\begin{align}
\left\vert 4\right\rangle  &  =\left\{  \frac{1}{4!}%
\begin{tabular}
[c]{|c|c|c|c|}\hline
$\mu_{1}^{1}$ & $\mu_{2}^{1}$ & $\mu_{3}^{1}$ & $\mu_{4}^{1}$\\\hline
\end{tabular}
\alpha_{-1}^{\mu_{1}^{1}}\alpha_{-1}^{\mu_{2}^{1}} \alpha_{-1}^{\mu_{3}^{1}%
}\alpha_{-1}^{\mu_{4}^{1}}+\frac{1}{2\cdot2!}%
\begin{tabular}
[c]{|c|c|}\hline
$\mu_{1}^{1}$ & $\mu_{2}^{1}$\\\hline
\end{tabular}
\otimes%
\begin{tabular}
[c]{|c|}\hline
$\mu_{1}^{2}$\\\hline
\end{tabular}
\alpha_{-1}^{\mu_{1}^{1}}\alpha_{-1}^{\mu_{2}^{1}}\alpha_{-2}^{\mu_{1}^{2}%
}\right.  \nn\\
&  \text{ }\left.  +\frac{1}{3}%
\begin{tabular}
[c]{|c|}\hline
$\mu_{1}^{1}$\\\hline
\end{tabular}
\otimes%
\begin{tabular}
[c]{|c|}\hline
$\mu_{1}^{3}$\\\hline
\end{tabular}
\alpha_{-1}^{\mu_{1}^{1}}\alpha_{-3}^{\mu_{1}^{3}}+\frac{1}{2^{2}\cdot2!}%
\begin{tabular}
[c]{|c|c|}\hline
$\mu_{1}^{2}$ & $\mu_{2}^{2}$\\\hline
\end{tabular}
\alpha_{-2}^{\mu_{1}^{2}}\alpha_{-2}^{\mu_{2}^{2}}+\frac{1}{4}%
\begin{tabular}
[c]{|c|}\hline
$\mu_{1}^{4}$\\\hline
\end{tabular}
\alpha_{-4}^{\mu_{1}^{4}}\right\}  \left\vert 0,k\right\rangle .
\end{align}
Next, we will apply the Virasoro constraints to the state Eq.(\ref{general 
state}
). The only Virasoro constraints which need to be considered are%
\begin{equation}
L_{1}\left\vert n\right\rangle =L_{2}\left\vert n\right\rangle =0,
\end{equation}
with $L_{m}$ the standard Virasoro operator%
\begin{equation}
L_{m}=\frac{1}{2}\sum_{n=-\infty}^{\infty}\alpha_{m+n}\cdot\alpha_{-n}.
\end{equation}
After taking care the symmetries of the Young tableaux, the Virasoro
constraints become
\begin{subequations}%
%

\begin{align}
L_{1}\left\vert n\right\rangle  &  =\sum_{m_{j}}\left[  k^{\mu_{1}^{1}%
}\right.  \overset{k}{\underset{j=1}{\otimes}}%
\begin{tabular}
[c]{|c|c|c|}\hline
$\mu_{1}^{j}$ & $\cdots$ & $\mu_{m_{j}}^{j}$\\\hline
\end{tabular}
\nonumber\\
&  +\sum_{i=2}^{m_{1}}%
\begin{tabular}
[c]{|c|c|c|c|c|}\hline
$\mu_{2}^{1}$ & $\cdots$ & $\hat{\mu}_{i}^{1}$ & $\cdots$ & $\mu_{m_{1}}^{1}%
$\\\hline
\end{tabular}
\otimes%
\begin{tabular}
[c]{|c|c|c|c|}\hline
$\mu_{i}^{1}$ & $\mu_{1}^{2}$ & $\cdots$ & $\mu_{m_{2}}^{2}$\\\hline
\end{tabular}
\overset{k}{\underset{j\neq1,2}{\otimes}}%
\begin{tabular}
[c]{|c|c|c|}\hline
$\mu_{1}^{j}$ & $\cdots$ & $\mu_{m_{j}}^{j}$\\\hline
\end{tabular}
\nonumber\\
&  +\sum_{l=3}^{k}\left(  l-1\right)
\begin{tabular}
[c]{|c|c|c|}\hline
$\mu_{2}^{1}$ & $\cdots$ & $\mu_{m_{1}}^{1}$\\\hline
\end{tabular}
\otimes\sum_{i=1}^{m_{l-1}}%
\begin{tabular}
[c]{|c|c|c|c|c|}\hline
$\mu_{1}^{l-1}$ & $\cdots$ & $\hat{\mu}_{i}^{l-1}$ & $\cdots$ & $\mu_{m_{l-1}%
}^{l-1}$\\\hline
\end{tabular}
\nonumber\\
&  \left.  \otimes%
\begin{tabular}
[c]{|c|c|c|c|}\hline
$\mu_{i}^{l-1}$ & $\mu_{1}^{l}$ & $\cdots$ & $\mu_{m_{l}}^{l}$\\\hline
\end{tabular}
\overset{k}{\underset{j\neq1,l,l-1}{\otimes}}%
\begin{tabular}
[c]{|c|c|c|}\hline
$\mu_{1}^{j}$ & $\cdots$ & $\mu_{m_{j}}^{j}$\\\hline
\end{tabular}
\right] \nonumber\\
&  \frac{1}{\left(  m_{1}-1\right)  !}\alpha_{-1}^{\mu_{2}^{1}\cdots\mu
_{m_{1}}^{1}}\prod_{j\neq1}^{k}\frac{1}{j^{m_{j}}m_{j}!}\alpha_{-j}^{\mu
_{1}^{j}\cdots\mu_{m_{j}}^{j}} |0,k \rangle =0, \label{L1}%
\end{align}
and
\begin{align}
L_{2}\left\vert n\right\rangle  &  =\sum_{m_{j}}\left[  \frac{1}{2}\eta
^{\mu_{1}^{1}\mu_{2}^{1}}\right.  \overset{k}{\underset{j=1}{\otimes}}%
\begin{tabular}
[c]{|c|c|c|}\hline
$\mu_{1}^{j}$ & $\cdots$ & $\mu_{m_{j}}^{j}$\\\hline
\end{tabular}
\nonumber\\
&  +
\begin{tabular}
[c]{|c|c|c|}\hline
$\mu_{3}^{1}$ & $\cdots$ & $\mu_{m_{1}}^{1}$\\\hline
\end{tabular}
\otimes%
\begin{tabular}
[c]{|c|c|c|}\hline
$\mu_{1}^{2}$ & $\cdots$ & $\mu_{m_{2}+1}^{2}$\\\hline
\end{tabular}
k^{\mu_{m_{2}+1}^{2}}\overset{k}{\underset{j\neq1,2}{\otimes}}%
\begin{tabular}
[c]{|c|c|c|}\hline
$\mu_{1}^{j}$ & $\cdots$ & $\mu_{m_{j}}^{j}$\\\hline
\end{tabular}
\nonumber\\
&  +\sum_{i=3}^{m_{1}}%
\begin{tabular}
[c]{|c|c|c|c|c|}\hline
$\mu_{3}^{1}$ & $\cdots$ & $\hat{\mu}_{i}^{1}$ & $\cdots$ & $\mu_{m_{1}}^{1}%
$\\\hline
\end{tabular}
\otimes%
\begin{tabular}
[c]{|c|c|c|c|}\hline
$\mu_{i}^{1}$ & $\mu_{1}^{3}$ & $\cdots$ & $\mu_{m_{3}}^{3}$\\\hline
\end{tabular}
\overset{k}{\underset{j\neq1,3}{\otimes}}%
\begin{tabular}
[c]{|c|c|c|}\hline
$\mu_{1}^{j}$ & $\cdots$ & $\mu_{m_{j}}^{j}$\\\hline
\end{tabular}
\nonumber\\
&  +\sum_{l=4}^{k}\left(  l-2\right)
\begin{tabular}
[c]{|c|c|c|}\hline
$\mu_{3}^{1}$ & $\cdots$ & $\mu_{m_{1}}^{1}$\\\hline
\end{tabular}
\otimes\sum_{i=1}^{m_{l-2}}%
\begin{tabular}
[c]{|c|c|c|c|c|}\hline
$\mu_{1}^{l-2}$ & $\cdots$ & $\hat{\mu}_{i}^{l-2}$ & $\cdots$ & $\mu_{m_{l}%
}^{l-2}$\\\hline
\end{tabular}
\nonumber\\
&  \left.  \overset{k}{\otimes%
\begin{tabular}
[c]{|c|c|c|c|}\hline
$\mu_{i}^{l-2}$ & $\mu_{1}^{l}$ & $\cdots$ & $\mu_{m_{l}}^{l}$\\\hline
\end{tabular}
\underset{j\neq1,l,l-2}{\otimes}}%
\begin{tabular}
[c]{|c|c|c|}\hline
$\mu_{1}^{j}$ & $\cdots$ & $\mu_{m_{j}}^{j}$\\\hline
\end{tabular}
\right] \nonumber\\
&  \frac{1}{\left(  m_{1}-2\right)  !}\alpha_{-1}^{\mu_{3}^{1}\cdots\mu
_{m_{1}}^{1}}\prod_{j\neq1}^{k}\alpha_{-j}^{\mu_{1}^{j}\cdots\mu_{m_{j}}^{j}}
| 0,k \rangle=0. \label{L2}%
\end{align}%
\end{subequations}%
A hat on an index means that the index is skipped there (and it should appear
somewhere else). In the above derivation we have used the identity for the
Young tableaux
\begin{align}%
\begin{tabular}
[c]{|c|c|c|}\hline
$1$ & $\cdots$ & $p$\\\hline
\end{tabular}
&  =\frac{1}{p}\left[  1+\sigma_{\left(  21\right)  }+\sigma_{\left(
321\right)  }+\cdots\sigma_{\left(  p\cdots1\right)  }\right]
\begin{tabular}
[c]{|c|c|c|}\hline
$2$ & $\cdots$ & $p$\\\hline
\end{tabular}
\otimes%
\begin{tabular}
[c]{|c|}\hline
$1$\\\hline
\end{tabular}
\nonumber\\
&  =\frac{1}{p}\sum_{i=1}^{p}\sigma_{\left(  i1\right)  }%
\begin{tabular}
[c]{|c|c|c|}\hline
$2$ & $\cdots$ & $p$\\\hline
\end{tabular}
\otimes%
\begin{tabular}
[c]{|c|}\hline
$1$\\\hline
\end{tabular}
,
\end{align}
where $\sigma_{\left(  i\cdots j \right)  }$ are permutation operators.

States which satisfy the Virasoro constraints are physical states. What we are
going to show in the following is that, in the high-energy limit, the Virasoro
constraints turn out to be strong enough to give the linear relationship among
the physical states. To take the high-energy limit in the above equations
(\ref{L1}) and (\ref{L2}), we replace the indices $\left(  \mu_{i},\nu
_{i}\right)  $ by $L$ or $T$, and%
\begin{equation}
k^{\mu_{i}}\rightarrow\hat{m}e^{L}\text{, } \quad\eta^{\mu_{1}\mu_{2}%
}\rightarrow e^{T}e^{T},
\end{equation}
where $\hat{m}$ is the mass operator. The Virasoro constraints at high energy
are derived in Appendix \ref{High Energy} as Eqs.(\ref{a}) and (\ref{b}). 
  To solve
the constraints, we need the following lemma to further simplify them.

\textbf{Lemma}

\begin{equation}%
\begin{tabular}
[c]{|l|l|l|}\hline
$T$ & $\cdots$ & $T$\\\hline
\end{tabular}
\ \underset{l_{1}}{\underbrace{%
\begin{tabular}
[c]{|l|l|l|}\hline
$L$ & $\cdots$ & $L$\\\hline
\end{tabular}
\ }}\otimes\underset{m_{2}-l_{2}}{\underbrace{%
\begin{tabular}
[c]{|l|l|l|}\hline
$T$ & $\cdots$ & $T$\\\hline
\end{tabular}
\ }}%
\begin{tabular}
[c]{|l|l|l|}\hline
$L$ & $\cdots$ & $L$\\\hline
\end{tabular}
\ \otimes\underset{\left\{  m_{j},j\geq3\right\}  }{\underbrace{%
\begin{tabular}
[c]{|lll|}\hline
& $\cdots$ & \\\hline
\end{tabular}
\ }}\equiv0, \label{lemma}%
\end{equation}
except for (i) $l_{2}=m_{2}$, $m_{j}=0$ for $j\geq3$ and (ii) $l_{1}=2m$.

This lemma is equivalent to part of the results of sec. 3, but will also be
proved in Appendix C.2 by applying Virasoro constraints. Finally, the Virasoro
constraints at high energy reduce to%
\begin{subequations}%
\begin{align}
&  \hat{m}\underset{n-2q-2-2m}{\underbrace{%
\begin{tabular}
[c]{|l|l|l|}\hline
$T$ & $\cdots$ & $T$\\\hline
\end{tabular}
\ \ }}\underset{2m+2}{\underbrace{%
\begin{tabular}
[c]{|l|l|l|}\hline
$L$ & $\cdots$ & $L$\\\hline
\end{tabular}
\ \ }}\otimes\underset{q}{\underbrace{%
\begin{tabular}
[c]{|l|l|l|}\hline
$L$ & $\cdots$ & $L$\\\hline
\end{tabular}
\ \ }}\nonumber\\
+  &  \left(  2m+1\right)  \underset{n-2q-2-2m}{\underbrace{%
\begin{tabular}
[c]{|l|l|l|}\hline
$T$ & $\cdots$ & $T$\\\hline
\end{tabular}
\ \ }}\underset{2m}{\underbrace{%
\begin{tabular}
[c]{|l|l|l|}\hline
$L$ & $\cdots$ & $L$\\\hline
\end{tabular}
\ \ }}\otimes\underset{q+1}{\underbrace{%
\begin{tabular}
[c]{|l|l|l|}\hline
$L$ & $\cdots$ & $L$\\\hline
\end{tabular}
\ \ }}=0,\label{L1-}\\
&  \hat{m}\underset{n-2q-2-2m}{\underbrace{%
\begin{tabular}
[c]{|l|l|l|}\hline
$T$ & $\cdots$ & $T$\\\hline
\end{tabular}
\ \ }}\underset{2m}{\underbrace{%
\begin{tabular}
[c]{|l|l|l|}\hline
$L$ & $\cdots$ & $L$\\\hline
\end{tabular}
\ \ }}\otimes\underset{q+1}{\underbrace{%
\begin{tabular}
[c]{|l|l|l|}\hline
$L$ & $\cdots$ & $L$\\\hline
\end{tabular}
\ \ }}\nonumber\\
+  &  \frac{1}{2}\underset{n-2q-2m}{\underbrace{%
\begin{tabular}
[c]{|l|l|l|}\hline
$T$ & $\cdots$ & $T$\\\hline
\end{tabular}
\ \ }}\underset{2m}{\underbrace{%
\begin{tabular}
[c]{|l|l|l|}\hline
$L$ & $\cdots$ & $L$\\\hline
\end{tabular}
\ \ }}\otimes\underset{q}{\underbrace{%
\begin{tabular}
[c]{|l|l|l|}\hline
$L$ & $\cdots$ & $L$\\\hline
\end{tabular}
\ \ }}=0, \label{L2-}%
\end{align}
where we have renamed $m_{2}\rightarrow q$ and $m_{1}\rightarrow n-2q$.%
\end{subequations}%

By mathematical recursion, Eq.(\ref{L1-}) leads to
\begin{subequations}%
\begin{equation}
\underset{n-2q-2m}{\underbrace{%
\begin{tabular}
[c]{|l|l|l|}\hline
$T$ & $\cdots$ & $T$\\\hline
\end{tabular}
\ }}\underset{2m}{\underbrace{%
\begin{tabular}
[c]{|l|l|l|}\hline
$L$ & $\cdots$ & $L$\\\hline
\end{tabular}
\ }}\otimes\underset{q}{\underbrace{%
\begin{tabular}
[c]{|l|l|l|}\hline
$L$ & $\cdots$ & $L$\\\hline
\end{tabular}
\ }}=\frac{\left(  2m-1\right)  !!\left(  -\hat{m}\right)  ^{q}}{\left(
2m+2q-1\right)  !!}\underset{n-2q-2m}{\underbrace{%
\begin{tabular}
[c]{|l|l|l|}\hline
$T$ & $\cdots$ & $T$\\\hline
\end{tabular}
\ }}\underset{2m+2q}{\underbrace{%
\begin{tabular}
[c]{|l|l|l|}\hline
$L$ & $\cdots$ & $L$\\\hline
\end{tabular}
\ }}, \label{L1--}%
\end{equation}
and similarly, Eq.(\ref{L2-}) leads to%
\begin{equation}
\underset{n-2q-2m}{\underbrace{%
\begin{tabular}
[c]{|l|l|l|}\hline
$T$ & $\cdots$ & $T$\\\hline
\end{tabular}
\ }}\underset{2m}{\underbrace{%
\begin{tabular}
[c]{|l|l|l|}\hline
$L$ & $\cdots$ & $L$\\\hline
\end{tabular}
\ }}\otimes\underset{q}{\underbrace{%
\begin{tabular}
[c]{|l|l|l|}\hline
$L$ & $\cdots$ & $L$\\\hline
\end{tabular}
\ }}=\left(  -\frac{1}{2\hat{m}}\right)  ^{q}\underset{n-2m}{\underbrace{%
\begin{tabular}
[c]{|l|l|l|}\hline
$T$ & $\cdots$ & $T$\\\hline
\end{tabular}
\ }}\underset{2m}{\underbrace{%
\begin{tabular}
[c]{|l|l|l|}\hline
$L$ & $\cdots$ & $L$\\\hline
\end{tabular}
\ }}. \label{L2--}%
\end{equation}%
\end{subequations}%
Combining equations (\ref{L1--}) and (\ref{L2--}), we get%
\begin{equation}
\underset{n-2q-2m}{\underbrace{%
\begin{tabular}
[c]{|l|l|l|}\hline
$T$ & $\cdots$ & $T$\\\hline
\end{tabular}
\ }}\underset{2m}{\underbrace{%
\begin{tabular}
[c]{|l|l|l|}\hline
$L$ & $\cdots$ & $L$\\\hline
\end{tabular}
\ }}\otimes\underset{q}{\underbrace{%
\begin{tabular}
[c]{|l|l|l|}\hline
$L$ & $\cdots$ & $L$\\\hline
\end{tabular}
\ }}=\left(  -\frac{1}{2\hat{m}}\right)  ^{q}\frac{\left(  2k-1\right)
!!}{4^{m}\left(  n-1\right)  ^{m}}\underset{n}{\underbrace{%
\begin{tabular}
[c]{|l|l|l|}\hline
$T$ & $\cdots$ & $T$\\\hline
\end{tabular}
\ }}. \label{ratio}%
\end{equation}
This is equivalent to Eq.(\ref{main}).

To get the ratio for the specific physical states, we make the Young tableaux
decomposition%
\begin{align}
&  \underset{n-2q-2m}{\underbrace{%
\begin{tabular}
[c]{|l|l|l|}\hline
$T$ & $\cdots$ & $T$\\\hline
\end{tabular}
\ \ }}\underset{2m}{\underbrace{%
\begin{tabular}
[c]{|l|l|l|}\hline
$L$ & $\cdots$ & $L$\\\hline
\end{tabular}
\ \ }}\otimes\underset{q}{\underbrace{%
\begin{tabular}
[c]{|l|l|l|}\hline
$L$ & $\cdots$ & $L$\\\hline
\end{tabular}
\ \ }}\nonumber\\
&  =\sum_{l=0}^{\min\left\{  n-2q-2m,q\right\}  }%
\begin{tabular}
[c]{|c|rccccccc}\hline
$T$ &  & $\cdots$ &  &  & \multicolumn{1}{|c}{$T$} & \multicolumn{1}{|c}{$L$}
& \multicolumn{1}{|c}{$\cdots$} & \multicolumn{1}{|c|}{$L$}\\\hline
$L$ & \multicolumn{1}{|c}{$\cdots$} & \multicolumn{1}{|c}{$L$} &
\multicolumn{1}{|c}{} &  &  &  &  & \\\cline{1-3}%
\end{tabular}
\ \ \cdot\left(  l!C_{q}^{l}C_{n-2q-2m}^{l}\right)  ,
\end{align}
where $C_{q}^{l}=\frac{q!}{l!(q-l)!}$ and we have $(n-2q-2m)$ $T$'s and
$(2m+q-l)$ $L$'s in the first column, $(l)$ $L$'s in the second column in the
second line of the above equation. Therefore, we obtain
\begin{align}
&
\begin{tabular}
[c]{|c|rccccccc}\hline
$T$ &  & $\cdots$ &  &  & \multicolumn{1}{|c}{$T$} & \multicolumn{1}{|c}{$L$}
& \multicolumn{1}{|c}{$\cdots$} & \multicolumn{1}{|c|}{$L$}\\\hline
$L$ & \multicolumn{1}{|c}{$\cdots$} & \multicolumn{1}{|c}{$L$} &
\multicolumn{1}{|c}{} &  &  &  &  & \\\cline{1-3}%
\end{tabular}
\ \ \cdot\left(  l!C_{q}^{l}C_{n-2q-2m}^{l}\right) \nonumber\\
&  =\frac{l!C_{q}^{l}C_{n-2q-2m}^{l}}{\sum_{l=0}^{\min\left\{
n-2q-2m,q\right\}  }l!C_{q}^{l}C_{n-2q-2m}^{l}}\left(  -\frac{1}{2\hat{m}%
}\right)  ^{q}\frac{\left(  2m-1\right)  !!}{4^{m}\left(  n-1\right)  ^{m}%
}\underset{n}{\underbrace{%
\begin{tabular}
[c]{|l|l|l|}\hline
$T$ & $\cdots$ & $T$\\\hline
\end{tabular}
\ \ }},
\end{align}
which is consistent with the ratios Eqs.(\ref{54}), (\ref{61}), and 
(\ref{68}) for
$m^{2}=4,6,8$ respectively.

\section{Saddle point approximation for stringy amplitudes}

\label{SaddlePoint}

In previous sections, we have identified the leading high-energy amplitudes
and derived the ratios among high-energy amplitudes for members of a family at
given mass levels, based on decoupling principle. While deductive arguments
help to clarify the underlying assumptions and solidify the validity of
decoupling principle, it is instructive to compare it with a different
approach, such as the saddle-point approximation \cite{CHL}. Therefore, we
shall perform direct calculations to check the results obtained above and make
comparisons between these two approaches.

In this section, we give a direct verification of the ratios among leading
high-energy amplitudes based on the saddle-point method. The four-point
amplitudes to be calculated consist of one massive tensor and three tachyons.
Since we have shown that in the high-energy limit the only relevant states are
those corresponding to
\begin{equation}
(\alpha_{-1}^{T})^{n-2m-2q} (\alpha_{-1}^{L})^{2m} (\alpha_{-2}^{L})^{q} |0,
k\rangle, \hspace{1cm} - k^{2} = 2 (n-1),
\end{equation}
we only need to calculate the following four-point amplitude
\begin{equation}
\mathcal{T}^{(n,2m,q)} \equiv\int\prod_{i=1}^{4} dx_{i} \langle V_{1}
V_{2}^{(n,2m,q)} V_{3} V_{4} \rangle,
\end{equation}
where
\begin{align}
V_{2}^{(n,2m,q)}  &  \equiv(\partial X^{T})^{n-2m-2q} (\partial X^{P})^{2m}
(\partial^{2} X^{P})^{q} e^{i k_{2} X_{2}},\\
V_{i}  &  \equiv e^{i k_{i} X_{i}}, \hspace{2cm} i = 1, 3, 4.
\end{align}
Notice that here for leading high-energy amplitudes we replace the
polarization $L$ by $P$.

Using either path-integral or operator formalism, after $SL(2,R)$ gauge
fixing, we obtain the $s-t$ channel contribution to the stringy amplitude at
tree level
\begin{align}
\mathcal{T}^{(n,2m,q)}  &  \Rightarrow\int_{0}^{1}dx x^{(1,2)}(1-x)^{(2,3)}%
\left[  \frac{e^{T}\cdot k_{1}}{x}-\frac{e^{T}\cdot k_{3}}{1-x}\right]
^{n-2m-2q}\nonumber\\
&  \cdot\left[  \frac{e^{P}\cdot k_{1}}{x}-\frac{e^{P}\cdot k_{3}}
{1-x}\right]  ^{2m}\left[  -\frac{e^{P}\cdot k_{1}}{x^{2}}-\frac{e^{P}\cdot
k_{3}}{(1-x)^{2}}\right]^{q}, \label{singletensor}%
\end{align}
where we have simplified the inner products among momenta by defining
$(1,2) \equiv k_1 \cdot k_2$.

In order to apply the saddle-point method, we need to rewrite the amplitude
above into the ``canonical form''. That is,
\begin{equation}
\mathcal{T}^{(n,2m,q)} (K) = \int_{0}^{1} d x \mbox{ } u(x) e^{- K f(x)},
\label{integral}%
\end{equation}
where
\begin{align}
K  &  \equiv- (1,2) \rightarrow\frac{s}{2} \rightarrow2 E^{2},\\
\tau &  \equiv-\frac{(2,3)}{(1,2)} \rightarrow- \frac{t}{s} \rightarrow
\sin^{2} \frac{\phi}{2},\\
f(x)  &  \equiv\ln x - \tau\ln(1-x),\\
u(x)  &  \equiv\left[  \frac{(1,2)}{\hat m} \right]  ^{2m+q} (1-x)^{-n +
2m+2q} (f^{\prime})^{2m} (f^{\prime\prime})^{q} (- e^{T} \cdot k_{3})^{n - 2m
-2q}.
\end{align}
The saddle-point for the integration of moduli, $x = x_{0}$, is defined by
\begin{equation}
f^{\prime}(x_{0}) = 0,
\end{equation}
and we have
\begin{equation}
x_{0} = \frac{1}{1 - \tau}, \hspace{1cm} 1 - x_{0} = - \frac{\tau}{1 - \tau},
\hspace{1cm} f^{\prime\prime}(x_{0}) = (1 - \tau)^{3} \tau^{-1}.
\end{equation}
{}From the definition of $u(x)$, it is easy to see that
\begin{equation}
u(x_{0}) = u^{\prime}(x_{0}) = .... = u^{(2 m -1)} (x_{0}) = 0,
\end{equation}
and
\begin{equation}
u^{(2m)} (x_{0}) = \left[  \frac{(1,2)}{\hat m} \right]  ^{2m+q} (1 -
x_{0})^{-n+2m+2q} (2m)! (f_{0}^{\prime\prime})^{2m+q} (- e^{T} \cdot k_{3})^{n
- 2m - 2q}.
\end{equation}

With these inputs, one can easily evaluate the Gaussian integral associated
with the four-point amplitudes, Eq.(\ref{integral}),
\begin{align}
&  \int_{0}^{1} dx \mbox{ } u(x) e^{- K f(x)}\nonumber\\
&  = \sqrt{\frac{2 \pi}{K f^{\prime\prime}_{0}}} e^{- K f_{0}} \left[
\frac{u_{0}^{(2m)}}{2^{m} \ m! \ (f^{\prime\prime}_{0})^{m} \ K^{m}} + O
(\frac{1}{K^{m+1}}) \right] \nonumber\\
&  = \sqrt{\frac{2 \pi}{K f^{\prime\prime}_{0}}} e^{- K f_{0}} \left[
(-1)^{n-q} \frac{2^{n-2m-q}(2m)!}{m! \ {\hat m}^{2m+q}} \ \tau^{-\frac{n}{2}}
(1-\tau)^{\frac{3n}{2}} E^{n} + O (E^{n-2}) \right]  . \label{leading}%
\end{align}
This result shows explicitly that with one tensor and three tachyons, the
energy and angle dependence for the high-energy four-point amplitudes only
depend on the level $n$, and we can solve for the ratios among high-energy
amplitudes within the same family,
\begin{align}
\lim_{E \rightarrow\infty} \frac{\mathcal{T}^{(n,2m,q)}}{\mathcal{T}
^{(n,0,0)}}  &  = \frac{(-1)^{q} (2m)!}{m! (2 \hat m)^{2m+q}}\\
&  = (-\frac{2m-1}{\hat m}) .... (-\frac{3}{\hat m}) (-\frac{1}{\hat m}) (-
\frac{1}{2 \hat m})^{m+q},
\end{align}
which is consistent with Eq.(\ref{main}).

We conclude this section with three remarks. Firstly, from the saddle-point
approach, it is easy to see why the product of $\alpha_{-1}^{P}$ oscillators
induce energy suppression. Their contribution to the stringy amplitude is
proportional to powers of $f^{\prime}(x_{0})$, which is zero in the leading
order calculation. Secondly, one can also understand why only even numbers of
$\alpha_{-1}^{P}$ oscillators will survive for high-energy amplitudes based on
the structure of Gaussian integral in Eq.(\ref{integral}). While for a vertex
operator containing $(2m +1)$ $\alpha_{-1}^{P}$'s, we have $u(x_{0}) =
u^{\prime}(x_{0}) = .... = u^{(2 m)} (x_{0}) = 0$, and the leading
contribution comes from $u^{(2 m+1)} (x_{0}) (x-x_{0})^{2m+1}$, this gives
zero since the odd-power moments of Gaussian integral vanish. Finally, for the
alert readers, since we only discuss the $s-t$ channel contribution to the
scattering amplitudes, the integration range for the $x$ variable seems to
devoid of a direct application of saddle-point method. Presumably, we can
apply the saddle-point approximation to the full amplitudes whose integration
range extends over whole real line. However, it is curious to see why $s-t$
channel alone has the same functional form as the full amplitude in the
high-energy limit. To see this, one can check that the leading contribution
Eq.(\ref{leading}) is actually ``form-invariant'' under any monotonous change
of variable $x= \xi(y), v(y) \equiv\xi^{\prime}(y) u(\xi(y)), g(y) \equiv
f(\xi(y))$. If the analytic structure of the integrand is no concern, we can
justify the use of saddle-point approximation even for the case of $s-t$ channel.

\section{Remarks}

\subsection{Virasoro generators at high energies}

In retrospect, the main result (\ref{main}) can be very easily obtained from
the Virasoro generators if we accept that we only need to consider states of
the form
\begin{equation}
\label{ourstates}\left(  \a^{T}_{-1}\right)  ^{n-m-2q} \left(  \a^{L}%
_{-1}\right)  ^{m} \left(  \a^{L}_{-2}\right)  ^{q} |0,k\rangle.
\end{equation}
On the space of these states, the Virasoro generators are effectively
\begin{align}
\tL_{-1}  &  = \hat{m} \a^{L}_{-1} + \a^{L}_{-2} \a^{L}_{1},\label{tL1}\\
\tL_{-2}  &  = \hat{m} \a^{L}_{-2} + \frac{1}{2}\a^{T}_{-1} a^{T}%
_{-1},\label{tL2}\\
\tL_{-n}  &  = 0 \quad\mbox{for} \quad n \geq3,
\end{align}
where we have also replaced $e^{P}$ by $e^{L}$.

Using these deformed Virasoro generators, we create high-energy approximations
of spurious states. On the back of an envelope, one can check that the
decoupling of the spurious states created by $\tL_{-1}$ and $\tL_{-2}$ implies
(\ref{1}) and (\ref{2}), respectively (with $P$ replaced by $L$). In short,
$\tL_{-1}$ tells us how to trade $\a^{L}_{-1}\a^{L}_{-1}$ for $\a^{L}_{-2}$,
and $\tL_{-2}$ tells us how to trade $\a^{L}_{-2}$ for $\a^{T}_{-1}\a^{T}%
_{-1}$.

Similarly, the Hermitian conjugates of $\tL_{1}$ and $\tL_{2}$ can be used to
derive the same result by demanding that they annihilate states in the
high-energy limit.

\subsection{2 dimensional string}

Although we have shown that there exist infinitely many linear relations among
4-point functions which uniquely fix their ratios in the high-energy limit, it
is not totally clear that there is a hidden symmetry responsible for it.
However, we would like to claim that these linear relations are indeed the
manifestation of the long-sought hidden symmetry of string theory, and that we
are on the right track of understanding the symmetry. To persuade the readers,
we test our claim on a toy model of string theory --the 2 dimensional string theory.

While the hidden symmetry of the 26 dimensional bosonic string theory is still
at large, the symmetry of the 2 dimensional string theory is much better
understood. It is known to be associated with the discrete states
\begin{equation}
\psi^{\pm}_{JM} \sim\Delta(J, M, -i\sqrt{2}X) \exp\left[  \sqrt{2}(iMX(0) +
(\pm J-1)\phi(0))\right]  .
\end{equation}
Half of them $\psi^{+}_{JM}$ generate the $w_{\infty}$ algebra \cite{Winfinity}
\begin{equation}
\int\frac{dz}{2\pi i} \psi^{+}_{J_{1} M_{1}} \psi^{+}_{J_{2} M_{2}} \sim(J_{2}
M_{1} - J_{1} M_{2} ) \psi^{+}_{(J_{1}+J_{2}-1)(M_{1}+M_{2})}. \label{walgebra}
\end{equation}
Let us now check whether the $w_{\infty}$ symmetry is generated by the high
energy limit of zero-norm states. In \cite{ChungLee}, explicit expression for
a class of zero-norm states was given
\begin{align}
\label{G+}G^{+}_{JM}  &  \sim(J-M)! \Delta(J, M, -i\sqrt{2}X) \exp\left[
\sqrt{2}(iMX + (J-1)\phi\right]  + (-1)^{2J} \sum_{j=1}^{J-M} (J-M-1)!\times
\nn\\
&  \times\int\frac{dz}{2\pi i} \mathcal{D}(J, M, -i\sqrt{2}X(z), j)
\exp\left[  \sqrt{2}(i(M+1)X(z) + (J-1)\phi(z) - X(0))\right]  .
\end{align}
The notation needs some explanation. Here $\Delta(J, M, -i\sqrt{2}X)$ is
defined by
\begin{equation}
\label{Delta}\Delta(J, M, -i\sqrt{2}X) = \left|
\begin{array}
[c]{cccc}%
S_{2J-1} & S_{2J-2} & \cdots & S_{J+M}\\
S_{2J-2} & S_{2J-3} & \cdots & S_{J+M-1}\\
\cdots & \cdots & \cdots & \cdots\\
S_{J+M} & S_{J+M-1} & \cdots & S_{2M+1}%
\end{array}
\right|  ,
\end{equation}
where
\begin{equation}
S_{k} = S_{k}\left(  \left\{  \frac{-i\sqrt{2}}{k!}\del^{k} X(0)\right\}
\right)  , \quad\mbox{and} \quad S_{k} = 0 \quad\mbox{if} \quad k < 0,
\end{equation}
and $S_{k}(\{a_{i}\})$'s denote the Schur polynomial defined by
\begin{equation}
\exp\left(  \sum_{k=1}^{\infty} a_{k} x^{k} \right)  = \sum_{k=0}^{\infty}
S_{k}(\{a_{i}\}) x^{k}.
\end{equation}
$\mathcal{D}(J, M, -i\sqrt{2}X(z), j)$ is defined by a similar expression as
Eq.(\ref{Delta}), but with the $j$-th row replaced by $\{ (-z)^{j}-1-2J,
(-z)^{j}-2J, \cdots, (-z)^{j-J-M-2} \}$.
It was shown \cite{ChungLee} that zero-norm states in Eq.(\ref{G+}) generate a
$w_{\infty}$ algebra.

In the high-energy limit, the factors $\del^{k} X^{A}$ are generically
proportional to a linear combination of the momenta of other vertices, so it
scales with energy $E$. Thus $\mathcal{D}(J, M, -i\sqrt{2}X, j)$ is subleading
to $\Delta(J, M, -i\sqrt{2}X)$. Ignoring the second term in Eq.(\ref{G+}) for
this reason, we see that these zero-norm states indeed approach to the
discrete states $\psi^{+}_{JM}$ above! 
Thus, the $w_{\infty}$ algebra generated by Eq.(\ref{G+}) is identified to
$w_{\infty}$ symmetry in Eq.(\ref{walgebra}).
This result strongly suggests that the
linear relations among correlation functions obtained from HZNS are indeed
related to the hidden symmetry also for the 26 dimensional strings. Although
we still do not know what is the symmetry group, or how it acts on states,
this work sheds new light on the road to finding the answers.

\section*{Acknowledgment}

The authors thank Jiunn-Wei Chen, Tohru Eguchi, Koji Hashimoto, Hiroyuki Hata,
Takeo Inami, Yeong-Chuan Kao, Yoichi Kazama, Yutaka Matsuo, Tamiaki Yoneya for
helpful discussions. This work is supported in part by the National Science
Council and the National Center for Theoretical Sciences, Taiwan, R.O.C.

\appendix

\setcounter{equation}{0} \renewcommand{\theequation}{A.\arabic{equation}}

\section{Kinematic variables and notations}

\label{appendix}

For the readers' convenience, we list the expressions of the kinematic
variables involved in the evaluation of a 4-point function in this appendix.
In Fig.1, we take the scattering plane to be the $X^{1}-X^{2}$ plane. The
momenta of the particles are
\begin{align}
k_{1}  &  = (\sqrt{p^{2} + m_{1}^{2}}, -p, 0),\\
k_{2}  &  = (\sqrt{p^{2} + m_{2}^{2}}, p, 0),\\
k_{3}  &  = (-\sqrt{q^{2} + m_{3}^{2}}, -q\cos\phi, -q\sin\phi),\\
k_{4}  &  = (-\sqrt{q^{2} + m_{4}^{2}}, q\cos\phi, q\sin\phi).
\end{align}
They satisfy $k_{i}^{2} = -m_{i}^{2}$. In the high-energy limit, the
Mandelstam variables are
\begin{align}
s  &  \equiv-(k_{1} + k_{2})^{2} = 4 E^{2} + \mathcal{O}(1/E^{2}),\\
t  &  \equiv-(k_{2} + k_{3})^{2} = -4 \left(  E^{2} - \frac{\sum_{i=1}^{4}
m_{i}^{2}}{4}\right)  \sin^{2} \frac{\phi}{2} + \mathcal{O}(1/E^{2}),\\
u  &  \equiv-(k_{1} + k_{3})^{2} = -4 \left(  E^{2} - \frac{\sum_{i=1}^{4}
m_{i}^{2}}{4}\right)  \cos^{2} \frac{\phi}{2} + \mathcal{O}(1/E^{2}),
\end{align}
where $E$ is related to $p$ and $q$ as
\begin{equation}
E^{2} = p^{2} + \frac{m_{1}^{2} + m_{2}^{2}}{2} = q^{2} + \frac{m_{3}^{2} +
m_{4}^{2}}{2}.
\end{equation}
The polarization bases for the 4 particles are
\begin{align}
e^{L}(1) = \frac{1}{m_{1}} (p, - \sqrt{p^{2} + m_{1}^{2}}, 0),  &  e^{T}(1) =
(0, 0, - 1),\\
e^{L}(2) = \frac{1}{m_{2}} (p, \sqrt{p^{2} + m_{2}^{2}}, 0),  &  e^{T}(2) =
(0, 0, 1),\\
e^{L}(3) = \frac{1}{m_{3}} (-q, - \sqrt{q^{2} + m_{3}^{2}} \cos\phi, -
\sqrt{q^{2} + m_{3}^{2}} \sin\phi),  &  e^{T}(3) = (0, - \sin\phi, \cos
\phi),\\
e^{L}(4) = \frac{1}{m_{4}} (-q, \sqrt{q^{2} + m_{4}}\cos\phi, \sqrt{q^{2} +
m_{4}^{2}} \sin\phi),  &  e^{T}(4) = (0, \sin\phi, -\cos\phi).
\end{align}

\setcounter{equation}{0} \renewcommand{\theequation}{B.\arabic{equation}}

\section{High energy zero-norm states}

In this subsection, we explicitly calculate high-energy zero-norm states
(HZNS) of some low-lying mass level. We will also show that the decoupling of
these HZNS is equivalent to the decoupling of those spurious states used in
the text to derive the desired linear relations. In the old covariant first
quantization spectrum of open bosonic string theory, the solutions of physical
state conditions include positive-norm propagating states and two types of
zero-norm states. The latter are \cite{GSW}
\begin{align}
\text{Type I}:L_{-1}\left\vert x\right\rangle ,  &  \text{ where }%
L_{1}\left\vert x\right\rangle =L_{2}\left\vert x\right\rangle =0,\text{
}L_{0}\left\vert x\right\rangle =0;\label{D.1}\\
\text{Type II}:(L_{-2}+\frac{3}{2}L_{-1}^{2})\left\vert \widetilde
{x}\right\rangle ,  &  \text{ where }L_{1}\left\vert \widetilde{x}%
\right\rangle =L_{2}\left\vert \widetilde{x}\right\rangle =0,\text{ }%
(L_{0}+1)\left\vert \widetilde{x}\right\rangle =0. \label{D.2}%
\end{align}
Based on a simplified calculation of higher mass level positive-norm states in
\cite{2} , some general solutions of zero-norm states of Eqs.(\ref{D.1}) and
(\ref{D.2}) at arbitrary mass level were calculated in \cite{3}.
Eqs.(\ref{D.1}) and (\ref{D.2}) can be derived from Kac determinant in
conformal field theory. While type I states have zero-norm at any spacetime
dimension, type II states have zero-norm \textit{only} at D=26.

The solutions of Eqs.(\ref{D.1}) and (\ref{D.2}) up to the mass level
$m^{2}=4$ are listed as follows \cite{3}:

1. $m^{2}=-k^{2}=0:$%
\begin{equation}
L_{-1\text{ }}\left\vert x\right\rangle =k\cdot\alpha_{-1}\left\vert
0,k\right\rangle ;\left\vert x\right\rangle =\left\vert 0,k\right\rangle
;\left\vert x\right\rangle =\left\vert 0,k\right\rangle . \label{D.3}%
\end{equation}

2. $m^{2}=-k^{2}=2:$%
\begin{equation}
(L_{-2}+\frac{3}{2}L_{-1}^{2})\left\vert \widetilde{x}\right\rangle =[\frac
{1}{2}\alpha_{-1}\cdot\alpha_{-1}+\frac{5}{2}k\cdot\alpha_{-2}+\frac{3}%
{2}(k\cdot\alpha_{-1})^{2}]\left\vert 0,k\right\rangle ;\left\vert
\widetilde{x}\right\rangle =\left\vert 0,k\right\rangle , \label{D.4}%
\end{equation}%
\begin{equation}
L_{-1}\left\vert x\right\rangle =[\theta\cdot\alpha_{-2}+(k\cdot\alpha
_{-1})(\theta\cdot\alpha_{-1})]\left\vert 0,k\right\rangle ;\left\vert
x\right\rangle =\theta\cdot\alpha_{-1}\left\vert 0,k\right\rangle ,\theta\cdot
k=0. \label{D.5}%
\end{equation}

3. $m^{2}=-k^{2}=4:$%
\begin{align}
(L_{-2}+\frac{3}{2}L_{-1}^{2})\left\vert \widetilde{x}\right\rangle  &
=\{4\theta\cdot\alpha_{-3}+\frac{1}{2}(\alpha_{-1}\cdot\alpha_{-1}%
)(\theta\cdot\alpha_{-1})+\frac{5}{2}(k\cdot\alpha_{-2})(\theta\cdot
\alpha_{-1})\nonumber\\[0.01in]
&  +\frac{3}{2}(k\cdot\alpha_{-1})^{2}(\theta\cdot\alpha_{-1})+3(k\cdot
\alpha_{-1})(\theta\cdot\alpha_{-2})\}\left\vert 0,k\right\rangle ;\nonumber\\
\left\vert \widetilde{x}\right\rangle  &  =\theta\cdot\alpha_{-1}\left\vert
0,k\right\rangle ,k\cdot\theta=0, \label{D.6}%
\end{align}%
\begin{align}
L_{-1}\left\vert x\right\rangle  &  =[2\theta_{\mu\nu}\alpha_{-1}^{\mu}%
\alpha_{-2}^{\nu}+k_{\lambda}\theta_{\mu\nu}\alpha_{-1}^{\lambda}\alpha
_{-1}^{\mu}\alpha_{-1}^{\nu}]\left\vert 0,k\right\rangle ;\nonumber\\
\left\vert x\right\rangle  &  =\theta_{\mu\nu}\alpha_{-1}^{\mu\nu}\left\vert
0,k\right\rangle ,k\cdot\theta=\eta^{\mu\nu}\theta_{\mu\nu}=0,\theta_{\mu\nu
}=\theta_{\nu\mu}, \label{D.7}%
\end{align}%
\begin{align}
L_{-1}\left\vert x\right\rangle  &  =[\frac{1}{2}(k\cdot\alpha_{-1}%
)^{2}(\theta\cdot\alpha_{-1})+2\theta\cdot\alpha_{-3}+\frac{3}{2}(k\cdot
\alpha_{-1})(\theta\cdot\alpha_{-2})\nonumber\\
&  +\frac{1}{2}(k\cdot\alpha_{-2})(\theta\cdot\alpha_{-1})]\left\vert
0,k\right\rangle ;\text{ \ }\nonumber\\
\text{\ }\left\vert x\right\rangle  &  =[2\theta\cdot\alpha_{-2}+(k\cdot
\alpha_{-1})(\theta\cdot\alpha_{-1})]\left\vert 0,k\right\rangle ,\theta\cdot
k=0, \label{D.8}%
\end{align}%
\begin{align}
L_{-1}\left\vert x\right\rangle  &  =[\frac{17}{4}(k\cdot\alpha_{-1}%
)^{3}+\frac{9}{2}(k\cdot\alpha_{-1})(\alpha_{-1}\cdot\alpha_{-1}%
)+9(\alpha_{-1}\cdot\alpha_{-2})\nonumber\\
&  +21(k\cdot\alpha_{-1})(k\cdot\alpha_{-2})+25(k\cdot\alpha_{-3})]\left\vert
0,k\right\rangle ;\nonumber\\
\left\vert x\right\rangle  &  =[\frac{25}{2}k\cdot\alpha_{-2}+\frac{9}%
{2}\alpha_{-1}\cdot\alpha_{-1}+\frac{17}{4}(k\cdot\alpha_{-1})^{2}]\left\vert
0,k\right\rangle . \label{D.9}%
\end{align}

Note that there are two degenerate vector zero-norm states, Eq.(\ref{D.6}) for
type II and Eq.(\ref{D.8}) for type I, at mass level $m^{2}=4$. For mass level
$m^{2}=2,$ the high energy limit of Eqs. (\ref{D.5}) and (\ref{D.4}) are
calculated to be
\begin{align}
L_{-1}(\theta\cdot\alpha_{-1})\left\vert 0\right\rangle  &  \rightarrow
\sqrt{2}\alpha_{-1}^{L}\alpha_{-1}^{L}+\alpha_{-2}^{L}\left\vert
0\right\rangle ;\label{D.10}\\
(L_{-2}+\frac{3}{2}L_{-1}^{2})\left\vert 0\right\rangle  &  \rightarrow
(\sqrt{2}\alpha_{-2}^{L}+\frac{1}{2}\alpha_{-1}^{T}\alpha_{-1}^{T})\left\vert
0\right\rangle \label{D.11}\\
&  +\frac{3}{2}(2\alpha_{-1}^{L}\alpha_{-1}^{L}+\sqrt{2}\alpha_{-2}%
^{L})\left\vert 0\right\rangle . \label{D.12}%
\end{align}
Note that Eq.(\ref{D.12}) is the high energy limit of the second term of type
II zero-norm state. It is easy to see that the decoupling of (\ref{D.10})
implies the decoupling of (\ref{D.12}). So one can neglect the effect of
(\ref{D.12}) even though it is of leading order in energy. It turns out that
this phenomena persists to any higher mass level as well. \textit{This
justifies that the decoupling of HZNS is equivalent to the decoupling of those
spurious states used in the text to derive the desired linear relations.} By
solving Eqs.(\ref{D.10}) and (\ref{D.11}), we get the desired linear relation,
$\mathcal{T}_{TT}:\mathcal{T}_{L}:\mathcal{T}_{LL}=4:-\sqrt{2}:1$.
\ Similarly, the high-energy limit of Eqs.(\ref{D.6})-(\ref{D.9}) are
calculated to be%
\begin{align}
(L_{-2}+\frac{3}{2}L_{-1}^{2})\left\vert 0\right\rangle  &  \rightarrow
(4\alpha_{-1}^{(T}\alpha_{-2}^{L)}+\frac{1}{2}\alpha_{-1}^{T}\alpha_{-1}%
^{T}\alpha_{-1}^{T})\left\vert 0\right\rangle \label{D.13}\\
&  +\frac{3}{2}(4\alpha_{-1}^{(L}\alpha_{-1}^{L}\alpha_{-1}^{T)}+4\alpha
_{-1}^{(T}\alpha_{-2}^{L)})\left\vert 0\right\rangle ;\label{D.14}\\
L_{-1}(\theta_{\mu\nu}\alpha_{-1}^{\mu\nu})\left\vert 0\right\rangle  &
\rightarrow\lbrack2\alpha_{-1}^{(T}\alpha_{-2}^{L)}+2\alpha_{-1}^{(L}%
\alpha_{-1}^{L}\alpha_{-1}^{T)}]\left\vert 0\right\rangle ;\label{D.15}\\
L_{-1}[2\theta\cdot\alpha_{-2}+(k\cdot\alpha_{-1})(\theta\cdot\alpha
_{-1})]\left\vert 0\right\rangle  &  \rightarrow(4\alpha_{-1}^{(L}\alpha
_{-1}^{L}\alpha_{-1}^{T)}+4\alpha_{-1}^{(T}\alpha_{-2}^{L)})\left\vert
0\right\rangle ;\label{D.16}\\
L_{-1}[\frac{25}{2}k\cdot\alpha_{-2}+\frac{9}{2}\alpha_{-1}\cdot\alpha
_{-1}+\frac{17}{4}(k\cdot\alpha_{-1})^{2}]\left\vert 0\right\rangle  &
\rightarrow0. \label{D.17}%
\end{align}

It is easy to see that the decoupling of Eq.(\ref{D.15}) or (\ref{D.16})
implies the decoupling of Eq.(\ref{D.14}). By solving the equations, one gets
$\mathcal{T}_{TTT}:\mathcal{T}_{LLT}:\mathcal{T}_{(LT)}:\mathcal{T}%
_{[LT]}=8:1:-1:1.$

\vskip .8cm \baselineskip 22pt \setcounter{equation}{0} \renewcommand{\theequation}{C.\arabic{equation}}

\section{Virasoro constraints}

\subsection{High energy limit of Virasoro constraints}

\label{High Energy}

To take the high-energy limit for the Virasoro constraints, we replace the
indices $\left(  \mu_{i},\nu_{i}\right)  $ by $L$ or $T$, and%
\begin{equation}
k^{\mu_{i}}\rightarrow\hat{m}e^{L}\text{, }\eta^{\mu_{1}\mu_{2}}\rightarrow
e^{T}e^{T}.
\end{equation}
Equations (\ref{L1}) and (\ref{L2}) become%
\begin{subequations}%
\begin{align}
0  &  =\hat{m}%
\begin{tabular}
[c]{|c|c|c|c|}\hline
$L$ & $\mu_{2}^{1}$ & $\cdots$ & $\mu_{m_{1}}^{1}$\\\hline
\end{tabular}
\overset{k}{\underset{j\neq1}{\otimes}}%
\begin{tabular}
[c]{|c|c|c|}\hline
$\mu_{1}^{j}$ & $\cdots$ & $\mu_{m_{j}}^{j}$\\\hline
\end{tabular}
\nonumber\\
&  +\sum_{i=2}^{m_{1}}%
\begin{tabular}
[c]{|c|c|c|c|c|}\hline
$\mu_{2}^{1}$ & $\cdots$ & $\hat{\mu}_{i}^{1}$ & $\cdots$ & $\mu_{m_{1}}^{1}%
$\\\hline
\end{tabular}
\otimes%
\begin{tabular}
[c]{|c|c|c|c|}\hline
$\mu_{i}^{1}$ & $\mu_{1}^{2}$ & $\cdots$ & $\mu_{m_{2}}^{2}$\\\hline
\end{tabular}
\overset{k}{\underset{j\neq1,2}{\otimes}}%
\begin{tabular}
[c]{|c|c|c|}\hline
$\mu_{1}^{j}$ & $\cdots$ & $\mu_{m_{j}}^{j}$\\\hline
\end{tabular}
\nonumber\\
&  +\sum_{l=3}^{k}\left(  l-1\right)
\begin{tabular}
[c]{|c|c|c|}\hline
$\mu_{2}^{1}$ & $\cdots$ & $\mu_{m_{1}}^{1}$\\\hline
\end{tabular}
\nonumber\\
&  \otimes\sum_{i=1}^{m_{l-1}}%
\begin{tabular}
[c]{|c|c|c|c|c|}\hline
$\mu_{1}^{l-1}$ & $\cdots$ & $\hat{\mu}_{i}^{l-1}$ & $\cdots$ & $\mu_{m_{l-1}%
}^{l-1}$\\\hline
\end{tabular}
\otimes%
\begin{tabular}
[c]{|c|c|c|c|}\hline
$\mu_{i}^{l-1}$ & $\mu_{1}^{l}$ & $\cdots$ & $\mu_{m_{l}}^{l}$\\\hline
\end{tabular}
\overset{k}{\underset{j\neq1,l,l-1}{\otimes}}%
\begin{tabular}
[c]{|c|c|c|}\hline
$\mu_{1}^{j}$ & $\cdots$ & $\mu_{m_{j}}^{j}$\\\hline
\end{tabular}
,
\end{align}
and%
\begin{align}
0  &  =\frac{1}{2}%
\begin{tabular}
[c]{|c|c|c|c|c|}\hline
$T$ & $T$ & $\mu_{3}^{1}$ & $\cdots$ & $\mu_{m_{1}}^{1}$\\\hline
\end{tabular}
\overset{k}{\underset{j\neq1}{\otimes}}%
\begin{tabular}
[c]{|c|c|c|}\hline
$\mu_{1}^{j}$ & $\cdots$ & $\mu_{m_{j}}^{j}$\\\hline
\end{tabular}
\nonumber\\
&  +\hat{m}%
\begin{tabular}
[c]{|c|c|c|}\hline
$\mu_{3}^{1}$ & $\cdots$ & $\mu_{m_{1}}^{1}$\\\hline
\end{tabular}
\otimes%
\begin{tabular}
[c]{|c|c|c|c|}\hline
$\mu_{1}^{2}$ & $\cdots$ & $\mu_{m_{2}}^{2}$ & $L$\\\hline
\end{tabular}
\overset{k}{\underset{j\neq1,2}{\otimes}}%
\begin{tabular}
[c]{|c|c|c|}\hline
$\mu_{1}^{j}$ & $\cdots$ & $\mu_{m_{j}}^{j}$\\\hline
\end{tabular}
\nonumber\\
&  +\sum_{i=3}^{m_{1}}%
\begin{tabular}
[c]{|c|c|c|c|c|}\hline
$\mu_{3}^{1}$ & $\cdots$ & $\hat{\mu}_{i}^{1}$ & $\cdots$ & $\mu_{m_{1}}^{1}%
$\\\hline
\end{tabular}
\otimes%
\begin{tabular}
[c]{|c|c|c|c|}\hline
$\mu_{i}^{1}$ & $\mu_{1}^{3}$ & $\cdots$ & $\mu_{m_{3}}^{3}$\\\hline
\end{tabular}
\overset{k}{\underset{j\neq1,3}{\otimes}}%
\begin{tabular}
[c]{|c|c|c|}\hline
$\mu_{1}^{j}$ & $\cdots$ & $\mu_{m_{j}}^{j}$\\\hline
\end{tabular}
\nonumber\\
&  +\sum_{l=4}^{k}\left(  l-2\right)
\begin{tabular}
[c]{|c|c|c|}\hline
$\mu_{3}^{1}$ & $\cdots$ & $\mu_{m_{1}}^{1}$\\\hline
\end{tabular}
\nonumber\\
&  \otimes\sum_{i=1}^{m_{l-2}}%
\begin{tabular}
[c]{|c|c|c|c|c|}\hline
$\mu_{1}^{l-2}$ & $\cdots$ & $\hat{\mu}_{i}^{l-2}$ & $\cdots$ & $\mu_{m_{l}%
}^{l-2}$\\\hline
\end{tabular}
\otimes%
\begin{tabular}
[c]{|c|c|c|c|}\hline
$\mu_{i}^{l-2}$ & $\mu_{1}^{l}$ & $\cdots$ & $\mu_{m_{l}}^{l}$\\\hline
\end{tabular}
\overset{k}{\underset{j\neq1,l,l-2}{\otimes}}%
\begin{tabular}
[c]{|c|c|c|}\hline
$\mu_{1}^{j}$ & $\cdots$ & $\mu_{m_{j}}^{j}$\\\hline
\end{tabular}
.
\end{align}%
\end{subequations}%
The indices and $\left\{  \mu_{i}^{j}\right\}  $ are symmetric and can be
chosen to have $l_{j}$ of $\left\{  L\right\}  $ which $0\leq l_{j}\leq m_{j}$
and $\left\{  T\right\}  $ for the rest. Thus%
\begin{subequations}
\begin{align}
0  &  =\hat{m}%
\begin{tabular}
[c]{|c|}\hline
$\mu_{2}^{1}$\\\hline
\end{tabular}
\underset{m_{1}-2-l_{1}}{\underbrace{%
\begin{tabular}
[c]{|l|l|l|}\hline
$T$ & $\cdots$ & $T$\\\hline
\end{tabular}
}}\underset{l_{1}+1}{\underbrace{%
\begin{tabular}
[c]{|l|l|l|}\hline
$L$ & $\cdots$ & $L$\\\hline
\end{tabular}
}}\overset{k}{\underset{j\neq1}{\otimes}}%
\begin{tabular}
[c]{|c|}\hline
$\mu_{1}^{j}$\\\hline
\end{tabular}
\underset{m_{j}-1-l_{j}}{\underbrace{%
\begin{tabular}
[c]{|l|l|l|}\hline
$T$ & $\cdots$ & $T$\\\hline
\end{tabular}
}}\underset{l_{j}}{\underbrace{%
\begin{tabular}
[c]{|l|l|l|}\hline
$L$ & $\cdots$ & $L$\\\hline
\end{tabular}
}}\nonumber\\
&  +\underset{m_{1}-2-l_{1}}{\underbrace{%
\begin{tabular}
[c]{|l|l|l|}\hline
$T$ & $\cdots$ & $T$\\\hline
\end{tabular}
}}\underset{l_{1}}{\underbrace{%
\begin{tabular}
[c]{|l|l|l|}\hline
$L$ & $\cdots$ & $L$\\\hline
\end{tabular}
}}\otimes%
\begin{tabular}
[c]{|c|c|}\hline
$\mu_{2}^{1}$ & $\mu_{1}^{2}$\\\hline
\end{tabular}
\underset{m_{2}-1-l_{2}}{\underbrace{%
\begin{tabular}
[c]{|l|l|l|}\hline
$T$ & $\cdots$ & $T$\\\hline
\end{tabular}
}}\underset{l_{2}}{\underbrace{%
\begin{tabular}
[c]{|l|l|l|}\hline
$L$ & $\cdots$ & $L$\\\hline
\end{tabular}
}}\nonumber\\
&  \overset{k}{\underset{j\neq1,2}{\otimes}}%
\begin{tabular}
[c]{|c|}\hline
$\mu_{1}^{j}$\\\hline
\end{tabular}
\underset{m_{j}-1-l_{j}}{\underbrace{%
\begin{tabular}
[c]{|l|l|l|}\hline
$T$ & $\cdots$ & $T$\\\hline
\end{tabular}
}}\underset{l_{j}}{\underbrace{%
\begin{tabular}
[c]{|l|l|l|}\hline
$L$ & $\cdots$ & $L$\\\hline
\end{tabular}
}}\nonumber\\
&  +\left(  m_{1}-2-l_{1}\right)
\begin{tabular}
[c]{|c|}\hline
$\mu_{2}^{1}$\\\hline
\end{tabular}
\underset{m_{1}-3-l_{1}}{\underbrace{%
\begin{tabular}
[c]{|l|l|l|}\hline
$T$ & $\cdots$ & $T$\\\hline
\end{tabular}
}}\underset{l_{1}}{\underbrace{%
\begin{tabular}
[c]{|l|l|l|}\hline
$L$ & $\cdots$ & $L$\\\hline
\end{tabular}
}}\otimes%
\begin{tabular}
[c]{|c|}\hline
$\mu_{1}^{2}$\\\hline
\end{tabular}
\underset{m_{2}-l_{2}}{\underbrace{%
\begin{tabular}
[c]{|l|l|l|}\hline
$T$ & $\cdots$ & $T$\\\hline
\end{tabular}
}}\underset{l_{2}}{\underbrace{%
\begin{tabular}
[c]{|l|l|l|}\hline
$L$ & $\cdots$ & $L$\\\hline
\end{tabular}
}}\nonumber\\
&  \overset{k}{\underset{j\neq1,2}{\otimes}}%
\begin{tabular}
[c]{|c|}\hline
$\mu_{1}^{j}$\\\hline
\end{tabular}
\underset{m_{j}-1-l_{j}}{\underbrace{%
\begin{tabular}
[c]{|l|l|l|}\hline
$T$ & $\cdots$ & $T$\\\hline
\end{tabular}
}}\underset{l_{j}}{\underbrace{%
\begin{tabular}
[c]{|l|l|l|}\hline
$L$ & $\cdots$ & $L$\\\hline
\end{tabular}
}}\nonumber\\
&  +l_{1}%
\begin{tabular}
[c]{|c|}\hline
$\mu_{2}^{1}$\\\hline
\end{tabular}
\underset{m_{1}-2-l_{1}}{\underbrace{%
\begin{tabular}
[c]{|l|l|l|}\hline
$T$ & $\cdots$ & $T$\\\hline
\end{tabular}
}}\underset{l_{1}-1}{\underbrace{%
\begin{tabular}
[c]{|l|l|l|}\hline
$L$ & $\cdots$ & $L$\\\hline
\end{tabular}
}}\otimes%
\begin{tabular}
[c]{|c|}\hline
$\mu_{1}^{2}$\\\hline
\end{tabular}
\underset{m_{2}-1-l_{2}}{\underbrace{%
\begin{tabular}
[c]{|l|l|l|}\hline
$T$ & $\cdots$ & $T$\\\hline
\end{tabular}
}}\underset{l_{2}+1}{\underbrace{%
\begin{tabular}
[c]{|l|l|l|}\hline
$L$ & $\cdots$ & $L$\\\hline
\end{tabular}
}}\nonumber\\
&  \overset{k}{\underset{j\neq1,2}{\otimes}}%
\begin{tabular}
[c]{|c|}\hline
$\mu_{1}^{j}$\\\hline
\end{tabular}
\underset{m_{j}-1-l_{j}}{\underbrace{%
\begin{tabular}
[c]{|l|l|l|}\hline
$T$ & $\cdots$ & $T$\\\hline
\end{tabular}
}}\underset{l_{j}}{\underbrace{%
\begin{tabular}
[c]{|l|l|l|}\hline
$L$ & $\cdots$ & $L$\\\hline
\end{tabular}
}}\nonumber\\
&  +\sum_{l=3}^{k}\left(  l-1\right)
\begin{tabular}
[c]{|c|}\hline
$\mu_{2}^{1}$\\\hline
\end{tabular}
\underset{m_{1}-2-l_{1}}{\underbrace{%
\begin{tabular}
[c]{|l|l|l|}\hline
$T$ & $\cdots$ & $T$\\\hline
\end{tabular}
}}\underset{l_{1}}{\underbrace{%
\begin{tabular}
[c]{|l|l|l|}\hline
$L$ & $\cdots$ & $L$\\\hline
\end{tabular}
}}\otimes\underset{m_{l-1}-1-l_{l-1}}{\underbrace{%
\begin{tabular}
[c]{|l|l|l|}\hline
$T$ & $\cdots$ & $T$\\\hline
\end{tabular}
}}\underset{l_{l-1}}{\underbrace{%
\begin{tabular}
[c]{|l|l|l|}\hline
$L$ & $\cdots$ & $L$\\\hline
\end{tabular}
}}\nonumber\\
&  \otimes%
\begin{tabular}
[c]{|c|c|}\hline
$\mu_{1}^{l-1}$ & $\mu_{1}^{l}$\\\hline
\end{tabular}
\underset{m_{l}-1-l_{l}}{\underbrace{%
\begin{tabular}
[c]{|l|l|l|}\hline
$T$ & $\cdots$ & $T$\\\hline
\end{tabular}
}}\underset{l_{l}}{\underbrace{%
\begin{tabular}
[c]{|l|l|l|}\hline
$L$ & $\cdots$ & $L$\\\hline
\end{tabular}
}}\overset{k}{\underset{j\neq1,l,l-1}{\otimes}}%
\begin{tabular}
[c]{|c|}\hline
$\mu_{1}^{j}$\\\hline
\end{tabular}
\underset{m_{j}-1-l_{j}}{\underbrace{%
\begin{tabular}
[c]{|l|l|l|}\hline
$T$ & $\cdots$ & $T$\\\hline
\end{tabular}
}}\underset{l_{j}}{\underbrace{%
\begin{tabular}
[c]{|l|l|l|}\hline
$L$ & $\cdots$ & $L$\\\hline
\end{tabular}
}}\nonumber\\
&  +\sum_{l=3}^{k}\left(  l-1\right)  \left(  m_{l-1}-1-l_{l-1}\right)
\begin{tabular}
[c]{|c|}\hline
$\mu_{2}^{1}$\\\hline
\end{tabular}
\underset{m_{1}-2-l_{1}}{\underbrace{%
\begin{tabular}
[c]{|l|l|l|}\hline
$T$ & $\cdots$ & $T$\\\hline
\end{tabular}
}}\underset{l_{1}}{\underbrace{%
\begin{tabular}
[c]{|l|l|l|}\hline
$L$ & $\cdots$ & $L$\\\hline
\end{tabular}
}}\otimes%
\begin{tabular}
[c]{|c|}\hline
$\mu_{1}^{l-1}$\\\hline
\end{tabular}
\underset{m_{l-1}-2-l_{l-1}}{\underbrace{%
\begin{tabular}
[c]{|l|l|l|}\hline
$T$ & $\cdots$ & $T$\\\hline
\end{tabular}
}}\underset{l_{l-1}}{\underbrace{%
\begin{tabular}
[c]{|l|l|l|}\hline
$L$ & $\cdots$ & $L$\\\hline
\end{tabular}
}}\nonumber\\
&  \otimes%
\begin{tabular}
[c]{|c|}\hline
$\mu_{1}^{l}$\\\hline
\end{tabular}
\underset{m_{l}-l_{l}}{\underbrace{%
\begin{tabular}
[c]{|l|l|l|}\hline
$T$ & $\cdots$ & $T$\\\hline
\end{tabular}
}}\underset{l_{l}}{\underbrace{%
\begin{tabular}
[c]{|l|l|l|}\hline
$L$ & $\cdots$ & $L$\\\hline
\end{tabular}
}}\overset{k}{\underset{j\neq1,l,l-1}{\otimes}}%
\begin{tabular}
[c]{|c|}\hline
$\mu_{1}^{j}$\\\hline
\end{tabular}
\underset{m_{j}-1-l_{j}}{\underbrace{%
\begin{tabular}
[c]{|l|l|l|}\hline
$T$ & $\cdots$ & $T$\\\hline
\end{tabular}
}}\underset{l_{j}}{\underbrace{%
\begin{tabular}
[c]{|l|l|l|}\hline
$L$ & $\cdots$ & $L$\\\hline
\end{tabular}
}}\nonumber\\
&  +\sum_{l=3}^{k}l_{l-1}\left(  l-1\right)
\begin{tabular}
[c]{|c|}\hline
$\mu_{2}^{1}$\\\hline
\end{tabular}
\underset{m_{1}-2-l_{1}}{\underbrace{%
\begin{tabular}
[c]{|l|l|l|}\hline
$T$ & $\cdots$ & $T$\\\hline
\end{tabular}
}}\underset{l_{1}}{\underbrace{%
\begin{tabular}
[c]{|l|l|l|}\hline
$L$ & $\cdots$ & $L$\\\hline
\end{tabular}
}}\otimes%
\begin{tabular}
[c]{|c|}\hline
$\mu_{1}^{l-1}$\\\hline
\end{tabular}
\underset{m_{l-1}-1-l_{l-1}}{\underbrace{%
\begin{tabular}
[c]{|l|l|l|}\hline
$T$ & $\cdots$ & $T$\\\hline
\end{tabular}
}}\underset{l_{l-1}-1}{\underbrace{%
\begin{tabular}
[c]{|l|l|l|}\hline
$L$ & $\cdots$ & $L$\\\hline
\end{tabular}
}}\nonumber\\
&  \otimes%
\begin{tabular}
[c]{|c|}\hline
$\mu_{1}^{l}$\\\hline
\end{tabular}
\underset{m_{l}-1-l_{l}}{\underbrace{%
\begin{tabular}
[c]{|l|l|l|}\hline
$T$ & $\cdots$ & $T$\\\hline
\end{tabular}
}}\underset{l_{l}+1}{\underbrace{%
\begin{tabular}
[c]{|l|l|l|}\hline
$L$ & $\cdots$ & $L$\\\hline
\end{tabular}
}}\overset{k}{\underset{j\neq1,l,l-1}{\otimes}}%
\begin{tabular}
[c]{|c|}\hline
$\mu_{1}^{j}$\\\hline
\end{tabular}
\underset{m_{j}-1-l_{j}}{\underbrace{%
\begin{tabular}
[c]{|l|l|l|}\hline
$T$ & $\cdots$ & $T$\\\hline
\end{tabular}
}}\underset{l_{j}}{\underbrace{%
\begin{tabular}
[c]{|l|l|l|}\hline
$L$ & $\cdots$ & $L$\\\hline
\end{tabular}
}},
\end{align}
and%
\begin{align}
0  &  =\frac{1}{2}%
\begin{tabular}
[c]{|c|}\hline
$\mu_{3}^{1}$\\\hline
\end{tabular}
\underset{m_{1}-1-l_{1}}{\underbrace{%
\begin{tabular}
[c]{|l|l|l|}\hline
$T$ & $\cdots$ & $T$\\\hline
\end{tabular}
}}\underset{l_{1}}{\underbrace{%
\begin{tabular}
[c]{|l|l|l|}\hline
$L$ & $\cdots$ & $L$\\\hline
\end{tabular}
}}\overset{k}{\underset{j\neq1}{\otimes}}%
\begin{tabular}
[c]{|c|}\hline
$\mu_{1}^{j}$\\\hline
\end{tabular}
\underset{m_{j}-1-l_{j}}{\underbrace{%
\begin{tabular}
[c]{|l|l|l|}\hline
$T$ & $\cdots$ & $T$\\\hline
\end{tabular}
}}\underset{l_{j}}{\underbrace{%
\begin{tabular}
[c]{|l|l|l|}\hline
$L$ & $\cdots$ & $L$\\\hline
\end{tabular}
}}\nonumber\\
&  +\hat{m}%
\begin{tabular}
[c]{|c|}\hline
$\mu_{3}^{1}$\\\hline
\end{tabular}
\underset{m_{1}-3-l_{1}}{\underbrace{%
\begin{tabular}
[c]{|l|l|l|}\hline
$T$ & $\cdots$ & $T$\\\hline
\end{tabular}
}}\underset{l_{1}}{\underbrace{%
\begin{tabular}
[c]{|l|l|l|}\hline
$L$ & $\cdots$ & $L$\\\hline
\end{tabular}
}}\otimes%
\begin{tabular}
[c]{|c|}\hline
$\mu_{1}^{2}$\\\hline
\end{tabular}
\underset{m_{2}-1-l_{2}}{\underbrace{%
\begin{tabular}
[c]{|l|l|l|}\hline
$T$ & $\cdots$ & $T$\\\hline
\end{tabular}
}}\underset{l_{2}+1}{\underbrace{%
\begin{tabular}
[c]{|l|l|l|}\hline
$L$ & $\cdots$ & $L$\\\hline
\end{tabular}
}}\nonumber\\
&  \overset{k}{\underset{j\neq1,2}{\otimes}}%
\begin{tabular}
[c]{|c|}\hline
$\mu_{1}^{j}$\\\hline
\end{tabular}
\underset{m_{j}-1-l_{j}}{\underbrace{%
\begin{tabular}
[c]{|l|l|l|}\hline
$T$ & $\cdots$ & $T$\\\hline
\end{tabular}
}}\underset{l_{j}}{\underbrace{%
\begin{tabular}
[c]{|l|l|l|}\hline
$L$ & $\cdots$ & $L$\\\hline
\end{tabular}
}}\nonumber\\
&  +\underset{m_{1}-3-l_{1}}{\underbrace{%
\begin{tabular}
[c]{|l|l|l|}\hline
$T$ & $\cdots$ & $T$\\\hline
\end{tabular}
}}\underset{l_{1}}{\underbrace{%
\begin{tabular}
[c]{|l|l|l|}\hline
$L$ & $\cdots$ & $L$\\\hline
\end{tabular}
}}\otimes%
\begin{tabular}
[c]{|c|c|}\hline
$\mu_{3}^{1}$ & $\mu_{1}^{3}$\\\hline
\end{tabular}
\underset{m_{3}-1-l_{3}}{\underbrace{%
\begin{tabular}
[c]{|l|l|l|}\hline
$T$ & $\cdots$ & $T$\\\hline
\end{tabular}
}}\underset{l_{3}}{\underbrace{%
\begin{tabular}
[c]{|l|l|l|}\hline
$L$ & $\cdots$ & $L$\\\hline
\end{tabular}
}}\nonumber\\
&  \overset{k}{\underset{j\neq1,3}{\otimes}}%
\begin{tabular}
[c]{|c|}\hline
$\mu_{1}^{j}$\\\hline
\end{tabular}
\underset{m_{j}-1-l_{j}}{\underbrace{%
\begin{tabular}
[c]{|l|l|l|}\hline
$T$ & $\cdots$ & $T$\\\hline
\end{tabular}
}}\underset{l_{j}}{\underbrace{%
\begin{tabular}
[c]{|l|l|l|}\hline
$L$ & $\cdots$ & $L$\\\hline
\end{tabular}
}}\nonumber\\
&  +\left(  m_{1}-3-l_{1}\right)
\begin{tabular}
[c]{|c|}\hline
$\mu_{3}^{1}$\\\hline
\end{tabular}
\underset{m_{1}-4-l_{1}}{\underbrace{%
\begin{tabular}
[c]{|l|l|l|}\hline
$T$ & $\cdots$ & $T$\\\hline
\end{tabular}
}}\underset{l_{1}}{\underbrace{%
\begin{tabular}
[c]{|l|l|l|}\hline
$L$ & $\cdots$ & $L$\\\hline
\end{tabular}
}}\otimes%
\begin{tabular}
[c]{|c|}\hline
$\mu_{1}^{3}$\\\hline
\end{tabular}
\underset{m_{3}-l_{3}}{\underbrace{%
\begin{tabular}
[c]{|l|l|l|}\hline
$T$ & $\cdots$ & $T$\\\hline
\end{tabular}
}}\underset{l_{3}}{\underbrace{%
\begin{tabular}
[c]{|l|l|l|}\hline
$L$ & $\cdots$ & $L$\\\hline
\end{tabular}
}}\nonumber\\
&  \overset{k}{\underset{j\neq1,3}{\otimes}}%
\begin{tabular}
[c]{|c|}\hline
$\mu_{1}^{j}$\\\hline
\end{tabular}
\underset{m_{j}-1-l_{j}}{\underbrace{%
\begin{tabular}
[c]{|l|l|l|}\hline
$T$ & $\cdots$ & $T$\\\hline
\end{tabular}
}}\underset{l_{j}}{\underbrace{%
\begin{tabular}
[c]{|l|l|l|}\hline
$L$ & $\cdots$ & $L$\\\hline
\end{tabular}
}}\nonumber\\
&  +l_{1}%
\begin{tabular}
[c]{|c|}\hline
$\mu_{3}^{1}$\\\hline
\end{tabular}
\underset{m_{1}-3-l_{1}}{\underbrace{%
\begin{tabular}
[c]{|l|l|l|}\hline
$T$ & $\cdots$ & $T$\\\hline
\end{tabular}
}}\underset{l_{1}-1}{\underbrace{%
\begin{tabular}
[c]{|l|l|l|}\hline
$L$ & $\cdots$ & $L$\\\hline
\end{tabular}
}}\otimes%
\begin{tabular}
[c]{|c|}\hline
$\mu_{1}^{3}$\\\hline
\end{tabular}
\underset{m_{3}-1-l_{3}}{\underbrace{%
\begin{tabular}
[c]{|l|l|l|}\hline
$T$ & $\cdots$ & $T$\\\hline
\end{tabular}
}}\underset{l_{3}+1}{\underbrace{%
\begin{tabular}
[c]{|l|l|l|}\hline
$L$ & $\cdots$ & $L$\\\hline
\end{tabular}
}}\nonumber\\
&  \overset{k}{\underset{j\neq1,3}{\otimes}}%
\begin{tabular}
[c]{|c|}\hline
$\mu_{1}^{j}$\\\hline
\end{tabular}
\underset{m_{j}-1-l_{j}}{\underbrace{%
\begin{tabular}
[c]{|l|l|l|}\hline
$T$ & $\cdots$ & $T$\\\hline
\end{tabular}
}}\underset{l_{j}}{\underbrace{%
\begin{tabular}
[c]{|l|l|l|}\hline
$L$ & $\cdots$ & $L$\\\hline
\end{tabular}
}}\nonumber\\
&  +\sum_{l=4}^{k}\left(  l-2\right)
\begin{tabular}
[c]{|c|}\hline
$\mu_{3}^{1}$\\\hline
\end{tabular}
\underset{m_{1}-3-l_{1}}{\underbrace{%
\begin{tabular}
[c]{|l|l|l|}\hline
$T$ & $\cdots$ & $T$\\\hline
\end{tabular}
}}\underset{l_{1}}{\underbrace{%
\begin{tabular}
[c]{|l|l|l|}\hline
$L$ & $\cdots$ & $L$\\\hline
\end{tabular}
}}\otimes\underset{m_{l-2}-1-l_{l-2}}{\underbrace{%
\begin{tabular}
[c]{|l|l|l|}\hline
$T$ & $\cdots$ & $T$\\\hline
\end{tabular}
}}\underset{l_{l-2}}{\underbrace{%
\begin{tabular}
[c]{|l|l|l|}\hline
$L$ & $\cdots$ & $L$\\\hline
\end{tabular}
}}\nonumber\\
&  \otimes%
\begin{tabular}
[c]{|c|c|}\hline
$\mu_{1}^{l-2}$ & $\mu_{1}^{l}$\\\hline
\end{tabular}
\underset{m_{l}-1-l_{l}}{\underbrace{%
\begin{tabular}
[c]{|l|l|l|}\hline
$T$ & $\cdots$ & $T$\\\hline
\end{tabular}
}}\underset{l_{l}}{\underbrace{%
\begin{tabular}
[c]{|l|l|l|}\hline
$L$ & $\cdots$ & $L$\\\hline
\end{tabular}
}}\overset{k}{\underset{j\neq1,l,l-2}{\otimes}}%
\begin{tabular}
[c]{|c|}\hline
$\mu_{1}^{j}$\\\hline
\end{tabular}
\underset{m_{j}-1-l_{j}}{\underbrace{%
\begin{tabular}
[c]{|l|l|l|}\hline
$T$ & $\cdots$ & $T$\\\hline
\end{tabular}
}}\underset{l_{j}}{\underbrace{%
\begin{tabular}
[c]{|l|l|l|}\hline
$L$ & $\cdots$ & $L$\\\hline
\end{tabular}
}}\nonumber\\
&  +\sum_{l=4}^{k}\left(  l-2\right)  \left(  m_{l-2}-1-l_{l-2}\right)
\begin{tabular}
[c]{|c|}\hline
$\mu_{3}^{1}$\\\hline
\end{tabular}
\underset{m_{1}-3-l_{1}}{\underbrace{%
\begin{tabular}
[c]{|l|l|l|}\hline
$T$ & $\cdots$ & $T$\\\hline
\end{tabular}
}}\underset{l_{1}}{\underbrace{%
\begin{tabular}
[c]{|l|l|l|}\hline
$L$ & $\cdots$ & $L$\\\hline
\end{tabular}
}}\otimes%
\begin{tabular}
[c]{|c|}\hline
$\mu_{1}^{l-2}$\\\hline
\end{tabular}
\underset{m_{l-2}-2-l_{l-2}}{\underbrace{%
\begin{tabular}
[c]{|l|l|l|}\hline
$T$ & $\cdots$ & $T$\\\hline
\end{tabular}
}}\underset{l_{l-2}}{\underbrace{%
\begin{tabular}
[c]{|l|l|l|}\hline
$L$ & $\cdots$ & $L$\\\hline
\end{tabular}
}}\nonumber\\
&  \otimes%
\begin{tabular}
[c]{|c|}\hline
$\mu_{1}^{l}$\\\hline
\end{tabular}
\underset{m_{l}-l_{l}}{\underbrace{%
\begin{tabular}
[c]{|l|l|l|}\hline
$T$ & $\cdots$ & $T$\\\hline
\end{tabular}
}}\underset{l_{l}}{\underbrace{%
\begin{tabular}
[c]{|l|l|l|}\hline
$L$ & $\cdots$ & $L$\\\hline
\end{tabular}
}}\overset{k}{\underset{j\neq1,l,l-2}{\otimes}}%
\begin{tabular}
[c]{|c|}\hline
$\mu_{1}^{j}$\\\hline
\end{tabular}
\underset{m_{j}-1-l_{j}}{\underbrace{%
\begin{tabular}
[c]{|l|l|l|}\hline
$T$ & $\cdots$ & $T$\\\hline
\end{tabular}
}}\underset{l_{j}}{\underbrace{%
\begin{tabular}
[c]{|l|l|l|}\hline
$L$ & $\cdots$ & $L$\\\hline
\end{tabular}
}}\nonumber\\
&  +\sum_{l=4}^{k}l_{l-2}\left(  l-2\right)
\begin{tabular}
[c]{|c|}\hline
$\mu_{3}^{1}$\\\hline
\end{tabular}
\underset{m_{1}-3-l_{1}}{\underbrace{%
\begin{tabular}
[c]{|l|l|l|}\hline
$T$ & $\cdots$ & $T$\\\hline
\end{tabular}
}}\underset{l_{1}}{\underbrace{%
\begin{tabular}
[c]{|l|l|l|}\hline
$L$ & $\cdots$ & $L$\\\hline
\end{tabular}
}}\otimes\sum_{i=2}^{m_{l-2}}%
\begin{tabular}
[c]{|c|}\hline
$\mu_{1}^{l-2}$\\\hline
\end{tabular}
\underset{m_{l-2}-1-l_{l-2}}{\underbrace{%
\begin{tabular}
[c]{|l|l|l|}\hline
$T$ & $\cdots$ & $T$\\\hline
\end{tabular}
}}\underset{l_{l-2}-1}{\underbrace{%
\begin{tabular}
[c]{|l|l|l|}\hline
$L$ & $\cdots$ & $L$\\\hline
\end{tabular}
}}\nonumber\\
&  \otimes%
\begin{tabular}
[c]{|c|}\hline
$\mu_{1}^{l}$\\\hline
\end{tabular}
\underset{m_{l}-1-l_{l}}{\underbrace{%
\begin{tabular}
[c]{|l|l|l|}\hline
$T$ & $\cdots$ & $T$\\\hline
\end{tabular}
}}\underset{l_{l}+1}{\underbrace{%
\begin{tabular}
[c]{|l|l|l|}\hline
$L$ & $\cdots$ & $L$\\\hline
\end{tabular}
}}\overset{k}{\underset{j\neq1,l,l-2}{\otimes}}%
\begin{tabular}
[c]{|c|}\hline
$\mu_{1}^{j}$\\\hline
\end{tabular}
\underset{m_{j}-1-l_{j}}{\underbrace{%
\begin{tabular}
[c]{|l|l|l|}\hline
$T$ & $\cdots$ & $T$\\\hline
\end{tabular}
}}\underset{l_{j}}{\underbrace{%
\begin{tabular}
[c]{|l|l|l|}\hline
$L$ & $\cdots$ & $L$\\\hline
\end{tabular}
}}.
\end{align}%
\end{subequations}%
There are still some undetermined parameters $\mu_{2}^{1}$, $\mu_{3}^{1}$ and
$\mu_{1}^{j}\left(  j\geq2\right)  $, which can be chosen to be $L$ or $T$, in
the above equations. However, it is easy to see that both choices lead to the
same equations. Therefore, we will set all of them to be $T$ in the following.
The final Virasoro constraints at high energy become%
\begin{subequations}%
\begin{align}
0  &  =\hat{m}\underset{m_{1}-1-l_{1}}{\underbrace{%
\begin{tabular}
[c]{|l|l|l|}\hline
$T$ & $\cdots$ & $T$\\\hline
\end{tabular}
}}\underset{l_{1}+1}{\underbrace{%
\begin{tabular}
[c]{|l|l|l|}\hline
$L$ & $\cdots$ & $L$\\\hline
\end{tabular}
}}\overset{k}{\underset{j\neq1}{\otimes}}\underset{m_{j}-l_{j}}{\underbrace{%
\begin{tabular}
[c]{|l|l|l|}\hline
$T$ & $\cdots$ & $T$\\\hline
\end{tabular}
}}\underset{l_{j}}{\underbrace{%
\begin{tabular}
[c]{|l|l|l|}\hline
$L$ & $\cdots$ & $L$\\\hline
\end{tabular}
}}\nonumber\\
&  +\left(  m_{1}-1-l_{1}\right)  \underset{m_{1}-2-l_{1}}{\underbrace{%
\begin{tabular}
[c]{|l|l|l|}\hline
$T$ & $\cdots$ & $T$\\\hline
\end{tabular}
}}\underset{l_{1}}{\underbrace{%
\begin{tabular}
[c]{|l|l|l|}\hline
$L$ & $\cdots$ & $L$\\\hline
\end{tabular}
}}\otimes\underset{m_{2}+1-l_{2}}{\underbrace{%
\begin{tabular}
[c]{|l|l|l|}\hline
$T$ & $\cdots$ & $T$\\\hline
\end{tabular}
}}\underset{l_{2}}{\underbrace{%
\begin{tabular}
[c]{|l|l|l|}\hline
$L$ & $\cdots$ & $L$\\\hline
\end{tabular}
}}\nonumber\\
&  \overset{k}{\underset{j\neq1,2}{\otimes}}\underset{m_{j}-l_{j}}%
{\underbrace{%
\begin{tabular}
[c]{|l|l|l|}\hline
$T$ & $\cdots$ & $T$\\\hline
\end{tabular}
}}\underset{l_{j}}{\underbrace{%
\begin{tabular}
[c]{|l|l|l|}\hline
$L$ & $\cdots$ & $L$\\\hline
\end{tabular}
}}\nonumber\\
&  +l_{1}\underset{m_{1}-1-l_{1}}{\underbrace{%
\begin{tabular}
[c]{|l|l|l|}\hline
$T$ & $\cdots$ & $T$\\\hline
\end{tabular}
}}\underset{l_{1}-1}{\underbrace{%
\begin{tabular}
[c]{|l|l|l|}\hline
$L$ & $\cdots$ & $L$\\\hline
\end{tabular}
}}\otimes\underset{m_{2}-l_{2}}{\underbrace{%
\begin{tabular}
[c]{|l|l|l|}\hline
$T$ & $\cdots$ & $T$\\\hline
\end{tabular}
}}\underset{l_{2}+1}{\underbrace{%
\begin{tabular}
[c]{|l|l|l|}\hline
$L$ & $\cdots$ & $L$\\\hline
\end{tabular}
}}\nonumber\\
&  \overset{k}{\underset{j\neq1,2}{\otimes}}\underset{m_{j}-l_{j}}%
{\underbrace{%
\begin{tabular}
[c]{|l|l|l|}\hline
$T$ & $\cdots$ & $T$\\\hline
\end{tabular}
}}\underset{l_{j}}{\underbrace{%
\begin{tabular}
[c]{|l|l|l|}\hline
$L$ & $\cdots$ & $L$\\\hline
\end{tabular}
}}\nonumber\\
&  +\sum_{l=3}^{k}\left(  l-1\right)  \left(  m_{l-1}-l_{l-1}\right)
\underset{m_{1}-1-l_{1}}{\underbrace{%
\begin{tabular}
[c]{|l|l|l|}\hline
$T$ & $\cdots$ & $T$\\\hline
\end{tabular}
}}\underset{l_{1}}{\underbrace{%
\begin{tabular}
[c]{|l|l|l|}\hline
$L$ & $\cdots$ & $L$\\\hline
\end{tabular}
}}\otimes\underset{m_{l-1}-1-l_{l-1}}{\underbrace{%
\begin{tabular}
[c]{|l|l|l|}\hline
$T$ & $\cdots$ & $T$\\\hline
\end{tabular}
}}\underset{l_{l-1}}{\underbrace{%
\begin{tabular}
[c]{|l|l|l|}\hline
$L$ & $\cdots$ & $L$\\\hline
\end{tabular}
}}\nonumber\\
&  \otimes\underset{m_{l}+1-l_{l}}{\underbrace{%
\begin{tabular}
[c]{|l|l|l|}\hline
$T$ & $\cdots$ & $T$\\\hline
\end{tabular}
}}\underset{l_{l}}{\underbrace{%
\begin{tabular}
[c]{|l|l|l|}\hline
$L$ & $\cdots$ & $L$\\\hline
\end{tabular}
}}\overset{k}{\underset{j\neq1,l,l-1}{\otimes}}\underset{m_{j}-l_{j}%
}{\underbrace{%
\begin{tabular}
[c]{|l|l|l|}\hline
$T$ & $\cdots$ & $T$\\\hline
\end{tabular}
}}\underset{l_{j}}{\underbrace{%
\begin{tabular}
[c]{|l|l|l|}\hline
$L$ & $\cdots$ & $L$\\\hline
\end{tabular}
}}\nonumber\\
&  +\sum_{l=3}^{k}l_{l-1}\left(  l-1\right)  \underset{m_{1}-1-l_{1}%
}{\underbrace{%
\begin{tabular}
[c]{|l|l|l|}\hline
$T$ & $\cdots$ & $T$\\\hline
\end{tabular}
}}\underset{l_{1}}{\underbrace{%
\begin{tabular}
[c]{|l|l|l|}\hline
$L$ & $\cdots$ & $L$\\\hline
\end{tabular}
}}\otimes\underset{m_{l-1}-l_{l-1}}{\underbrace{%
\begin{tabular}
[c]{|l|l|l|}\hline
$T$ & $\cdots$ & $T$\\\hline
\end{tabular}
}}\underset{l_{l-1}-1}{\underbrace{%
\begin{tabular}
[c]{|l|l|l|}\hline
$L$ & $\cdots$ & $L$\\\hline
\end{tabular}
}}\nonumber\\
&  \otimes\underset{m_{l}-l_{l}}{\underbrace{%
\begin{tabular}
[c]{|l|l|l|}\hline
$T$ & $\cdots$ & $T$\\\hline
\end{tabular}
}}\underset{l_{l}+1}{\underbrace{%
\begin{tabular}
[c]{|l|l|l|}\hline
$L$ & $\cdots$ & $L$\\\hline
\end{tabular}
}} \overset{k}{\underset{j\neq1,l,l-1}{\otimes}}\underset{m_{j}-l_{j}%
}{\underbrace{%
\begin{tabular}
[c]{|l|l|l|}\hline
$T$ & $\cdots$ & $T$\\\hline
\end{tabular}
}}\underset{l_{j}}{\underbrace{%
\begin{tabular}
[c]{|l|l|l|}\hline
$L$ & $\cdots$ & $L$\\\hline
\end{tabular}
}}, \label{a}%
\end{align}
and%

\begin{align}
0  &  =\frac{1}{2}\underset{m_{1}-l_{1}}{\underbrace{%
\begin{tabular}
[c]{|l|l|l|}\hline
$T$ & $\cdots$ & $T$\\\hline
\end{tabular}
}}\underset{l_{1}}{\underbrace{%
\begin{tabular}
[c]{|l|l|l|}\hline
$L$ & $\cdots$ & $L$\\\hline
\end{tabular}
}}\overset{k}{\underset{j\neq1}{\otimes}}\underset{m_{j}-l_{j}}{\underbrace{%
\begin{tabular}
[c]{|l|l|l|}\hline
$T$ & $\cdots$ & $T$\\\hline
\end{tabular}
}}\underset{l_{j}}{\underbrace{%
\begin{tabular}
[c]{|l|l|l|}\hline
$L$ & $\cdots$ & $L$\\\hline
\end{tabular}
}}\nonumber\\
&  +\hat{m}\underset{m_{1}-2-l_{1}}{\underbrace{%
\begin{tabular}
[c]{|l|l|l|}\hline
$T$ & $\cdots$ & $T$\\\hline
\end{tabular}
}}\underset{l_{1}}{\underbrace{%
\begin{tabular}
[c]{|l|l|l|}\hline
$L$ & $\cdots$ & $L$\\\hline
\end{tabular}
}}\otimes\underset{m_{2}-l_{2}}{\underbrace{%
\begin{tabular}
[c]{|l|l|l|}\hline
$T$ & $\cdots$ & $T$\\\hline
\end{tabular}
}}\underset{l_{2}+1}{\underbrace{%
\begin{tabular}
[c]{|l|l|l|}\hline
$L$ & $\cdots$ & $L$\\\hline
\end{tabular}
}}\nonumber\\
&  \overset{k}{\underset{j\neq1,2}{\otimes}}\underset{m_{j}-l_{j}}%
{\underbrace{%
\begin{tabular}
[c]{|l|l|l|}\hline
$T$ & $\cdots$ & $T$\\\hline
\end{tabular}
}}\underset{l_{j}}{\underbrace{%
\begin{tabular}
[c]{|l|l|l|}\hline
$L$ & $\cdots$ & $L$\\\hline
\end{tabular}
}}\nonumber\\
&  +\left(  m_{1}-2-l_{1}\right)  \underset{m_{1}-3-l_{1}}{\underbrace{%
\begin{tabular}
[c]{|l|l|l|}\hline
$T$ & $\cdots$ & $T$\\\hline
\end{tabular}
}}\underset{l_{1}}{\underbrace{%
\begin{tabular}
[c]{|l|l|l|}\hline
$L$ & $\cdots$ & $L$\\\hline
\end{tabular}
}}\otimes\underset{m_{3}+1-l_{3}}{\underbrace{%
\begin{tabular}
[c]{|l|l|l|}\hline
$T$ & $\cdots$ & $T$\\\hline
\end{tabular}
}}\underset{l_{3}}{\underbrace{%
\begin{tabular}
[c]{|l|l|l|}\hline
$L$ & $\cdots$ & $L$\\\hline
\end{tabular}
}}\nonumber\\
&  \overset{k}{\underset{j\neq1,3}{\otimes}}\underset{m_{j}-l_{j}}%
{\underbrace{%
\begin{tabular}
[c]{|l|l|l|}\hline
$T$ & $\cdots$ & $T$\\\hline
\end{tabular}
}}\underset{l_{j}}{\underbrace{%
\begin{tabular}
[c]{|l|l|l|}\hline
$L$ & $\cdots$ & $L$\\\hline
\end{tabular}
}}\nonumber\\
&  +l_{1}\underset{m_{1}-2-l_{1}}{\underbrace{%
\begin{tabular}
[c]{|l|l|l|}\hline
$T$ & $\cdots$ & $T$\\\hline
\end{tabular}
}}\underset{l_{1}-1}{\underbrace{%
\begin{tabular}
[c]{|l|l|l|}\hline
$L$ & $\cdots$ & $L$\\\hline
\end{tabular}
}}\otimes\underset{m_{3}-l_{3}}{\underbrace{%
\begin{tabular}
[c]{|l|l|l|}\hline
$T$ & $\cdots$ & $T$\\\hline
\end{tabular}
}}\underset{l_{3}+1}{\underbrace{%
\begin{tabular}
[c]{|l|l|l|}\hline
$L$ & $\cdots$ & $L$\\\hline
\end{tabular}
}}\nonumber\\
&  \overset{k}{\underset{j\neq1,3}{\otimes}}\underset{m_{j}-l_{j}}%
{\underbrace{%
\begin{tabular}
[c]{|l|l|l|}\hline
$T$ & $\cdots$ & $T$\\\hline
\end{tabular}
}}\underset{l_{j}}{\underbrace{%
\begin{tabular}
[c]{|l|l|l|}\hline
$L$ & $\cdots$ & $L$\\\hline
\end{tabular}
}}\nonumber\\
&  +\sum_{l=4}^{k}\left(  l-2\right)  \left(  m_{l-2}-l_{l-2}\right)
\underset{m_{1}-2-l_{1}}{\underbrace{%
\begin{tabular}
[c]{|l|l|l|}\hline
$T$ & $\cdots$ & $T$\\\hline
\end{tabular}
}}\underset{l_{1}}{\underbrace{%
\begin{tabular}
[c]{|l|l|l|}\hline
$L$ & $\cdots$ & $L$\\\hline
\end{tabular}
}}\otimes\underset{m_{l-2}-1-l_{l-2}}{\underbrace{%
\begin{tabular}
[c]{|l|l|l|}\hline
$T$ & $\cdots$ & $T$\\\hline
\end{tabular}
}}\underset{l_{l-2}}{\underbrace{%
\begin{tabular}
[c]{|l|l|l|}\hline
$L$ & $\cdots$ & $L$\\\hline
\end{tabular}
}}\nonumber\\
&  \otimes\underset{m_{l}+1-l_{l}}{\underbrace{%
\begin{tabular}
[c]{|l|l|l|}\hline
$T$ & $\cdots$ & $T$\\\hline
\end{tabular}
}}\underset{l_{l}}{\underbrace{%
\begin{tabular}
[c]{|l|l|l|}\hline
$L$ & $\cdots$ & $L$\\\hline
\end{tabular}
}} \overset{k}{\underset{j\neq1,l,l-2}{\otimes}}\underset{m_{j}-l_{j}%
}{\underbrace{%
\begin{tabular}
[c]{|l|l|l|}\hline
$T$ & $\cdots$ & $T$\\\hline
\end{tabular}
}}\underset{l_{j}}{\underbrace{%
\begin{tabular}
[c]{|l|l|l|}\hline
$L$ & $\cdots$ & $L$\\\hline
\end{tabular}
}}\nonumber\\
&  +\sum_{l=4}^{k}l_{l-2}\left(  l-2\right)  \underset{m_{1}-2-l_{1}%
}{\underbrace{%
\begin{tabular}
[c]{|l|l|l|}\hline
$T$ & $\cdots$ & $T$\\\hline
\end{tabular}
}}\underset{l_{1}}{\underbrace{%
\begin{tabular}
[c]{|l|l|l|}\hline
$L$ & $\cdots$ & $L$\\\hline
\end{tabular}
}}\otimes\sum_{i=2}^{m_{l-2}}\underset{m_{l-2}-l_{l-2}}{\underbrace{%
\begin{tabular}
[c]{|l|l|l|}\hline
$T$ & $\cdots$ & $T$\\\hline
\end{tabular}
}}\underset{l_{l-2}-1}{\underbrace{%
\begin{tabular}
[c]{|l|l|l|}\hline
$L$ & $\cdots$ & $L$\\\hline
\end{tabular}
}}\nonumber\\
&  \otimes\underset{m_{l}-l_{l}}{\underbrace{%
\begin{tabular}
[c]{|l|l|l|}\hline
$T$ & $\cdots$ & $T$\\\hline
\end{tabular}
}}\underset{l_{l}+1}{\underbrace{%
\begin{tabular}
[c]{|l|l|l|}\hline
$L$ & $\cdots$ & $L$\\\hline
\end{tabular}
}}\overset{k}{\underset{j\neq1,l,l-2}{\otimes}}\underset{m_{j}-l_{j}%
}{\underbrace{%
\begin{tabular}
[c]{|l|l|l|}\hline
$T$ & $\cdots$ & $T$\\\hline
\end{tabular}
}}\underset{l_{j}}{\underbrace{%
\begin{tabular}
[c]{|l|l|l|}\hline
$L$ & $\cdots$ & $L$\\\hline
\end{tabular}
}}. \label{b}%
\end{align}%
\end{subequations}%

\subsection{Proof of the lemma (\ref{lemma})}

In this subsection, we prove the lemma given in Sec.4.2 as follows%

\begin{equation}%
\begin{tabular}
[c]{|l|l|l|}\hline
$T$ & $\cdots$ & $T$\\\hline
\end{tabular}
\ \underset{l_{1}}{\underbrace{%
\begin{tabular}
[c]{|l|l|l|}\hline
$L$ & $\cdots$ & $L$\\\hline
\end{tabular}
\ }}\otimes\underset{m_{2}-l_{2}}{\underbrace{%
\begin{tabular}
[c]{|l|l|l|}\hline
$T$ & $\cdots$ & $T$\\\hline
\end{tabular}
\ }}%
\begin{tabular}
[c]{|l|l|l|}\hline
$L$ & $\cdots$ & $L$\\\hline
\end{tabular}
\ \otimes\underset{\left\{  m_{j},j\geq3\right\}  }{\underbrace{%
\begin{tabular}
[c]{|lll|}\hline
& $\cdots$ & \\\hline
\end{tabular}
\ }}\equiv0,
\end{equation}
except for (i) $l_{2}=m_{2}$, $m_{j}=0$ for $j\geq3$ and (ii) $l_{1}=2m$.

\begin{proof}
In the high-energy limit, we only need to consider the leading energy terms.
To count the energy scaling behavior, the rule is the same as Eq.(20): each
$T$ contributes a factor of energy $E$ and each $L$ contributes $E^{2}$. Any
terms with total energy order level less than $n$ are sub-leading terms and
can be ignored.

(i) If $l_{2}\neq m_{2}$ and $m_{j}\neq0,j\geq3$, then in equation (\ref{a}),

\begin{enumerate}
\item for $l_{1}=0$, all terms except the first term are sub-leading, then%
\begin{equation}%
\begin{tabular}
[c]{|l|l|l|}\hline
$T$ & $\cdots$ & $T$\\\hline
\end{tabular}
\underset{1}{\underbrace{%
\begin{tabular}
[c]{|l|l|l|}\hline
$L$ & $\cdots$ & $L$\\\hline
\end{tabular}
}}\overset{k}{\underset{j\neq1}{\otimes}}\underset{m_{j}-l_{j}}{\underbrace{%
\begin{tabular}
[c]{|l|l|l|}\hline
$T$ & $\cdots$ & $T$\\\hline
\end{tabular}
}}\underset{l_{j}}{\underbrace{%
\begin{tabular}
[c]{|l|l|l|}\hline
$L$ & $\cdots$ & $L$\\\hline
\end{tabular}
}}=0, \label{l=0}%
\end{equation}

\item for $l_{1}=1$, the third term is sub-leading, and (\ref{l=0}) implies
all other terms except the first term are vanished, then%
\begin{equation}%
\begin{tabular}
[c]{|l|l|l|}\hline
$T$ & $\cdots$ & $T$\\\hline
\end{tabular}
\underset{2}{\underbrace{%
\begin{tabular}
[c]{|l|l|l|}\hline
$L$ & $\cdots$ & $L$\\\hline
\end{tabular}
}}\overset{k}{\underset{j\neq1}{\otimes}}\underset{m_{j}-l_{j}}{\underbrace{%
\begin{tabular}
[c]{|l|l|l|}\hline
$T$ & $\cdots$ & $T$\\\hline
\end{tabular}
}}\underset{l_{j}}{\underbrace{%
\begin{tabular}
[c]{|l|l|l|}\hline
$L$ & $\cdots$ & $L$\\\hline
\end{tabular}
}}=0,
\end{equation}

\item if for $l_{1}=l^{\prime}$,%
\begin{equation}%
\begin{tabular}
[c]{|l|l|l|}\hline
$T$ & $\cdots$ & $T$\\\hline
\end{tabular}
\underset{l^{\prime}-1}{\underbrace{%
\begin{tabular}
[c]{|l|l|l|}\hline
$L$ & $\cdots$ & $L$\\\hline
\end{tabular}
}}\overset{k}{\underset{j\neq1}{\otimes}}\underset{m_{j}-l_{j}}{\underbrace{%
\begin{tabular}
[c]{|l|l|l|}\hline
$T$ & $\cdots$ & $T$\\\hline
\end{tabular}
}}\underset{l_{j}}{\underbrace{%
\begin{tabular}
[c]{|l|l|l|}\hline
$L$ & $\cdots$ & $L$\\\hline
\end{tabular}
}}=0,
\end{equation}
and%
\begin{equation}%
\begin{tabular}
[c]{|l|l|l|}\hline
$T$ & $\cdots$ & $T$\\\hline
\end{tabular}
\underset{l^{\prime}}{\underbrace{%
\begin{tabular}
[c]{|l|l|l|}\hline
$L$ & $\cdots$ & $L$\\\hline
\end{tabular}
}}\overset{k}{\underset{j\neq1}{\otimes}}\underset{m_{j}-l_{j}}{\underbrace{%
\begin{tabular}
[c]{|l|l|l|}\hline
$T$ & $\cdots$ & $T$\\\hline
\end{tabular}
}}\underset{l_{j}}{\underbrace{%
\begin{tabular}
[c]{|l|l|l|}\hline
$L$ & $\cdots$ & $L$\\\hline
\end{tabular}
}}=0,
\end{equation}
(\ref{a}) implies all terms except the first term are vanished, then%
\begin{equation}%
\begin{tabular}
[c]{|l|l|l|}\hline
$T$ & $\cdots$ & $T$\\\hline
\end{tabular}
\underset{l^{\prime}+1}{\underbrace{%
\begin{tabular}
[c]{|l|l|l|}\hline
$L$ & $\cdots$ & $L$\\\hline
\end{tabular}
}}\overset{k}{\underset{j\neq1}{\otimes}}\underset{m_{j}-l_{j}}{\underbrace{%
\begin{tabular}
[c]{|l|l|l|}\hline
$T$ & $\cdots$ & $T$\\\hline
\end{tabular}
}}\underset{l_{j}}{\underbrace{%
\begin{tabular}
[c]{|l|l|l|}\hline
$L$ & $\cdots$ & $L$\\\hline
\end{tabular}
}}=0.
\end{equation}

\end{enumerate}

(ii) If $l_{2}=m_{2}$ and $m_{j}=0$ for $j\geq3$, then equation (\ref{a})
reduces to%
\begin{align}
\hat{m}\underset{m_{1}-1-l_{1}}{\underbrace{%
\begin{tabular}
[c]{|l|l|l|}\hline
$T$ & $\cdots$ & $T$\\\hline
\end{tabular}
}}\underset{l_{1}+1}{\underbrace{%
\begin{tabular}
[c]{|l|l|l|}\hline
$L$ & $\cdots$ & $L$\\\hline
\end{tabular}
}}  &  \otimes\underset{m_{2}}{\underbrace{%
\begin{tabular}
[c]{|l|l|l|}\hline
$L$ & $\cdots$ & $L$\\\hline
\end{tabular}
}}\nonumber\\
+l_{1}\underset{m_{1}-1-l_{1}}{\underbrace{%
\begin{tabular}
[c]{|l|l|l|}\hline
$T$ & $\cdots$ & $T$\\\hline
\end{tabular}
}}\underset{l_{1}-1}{\underbrace{%
\begin{tabular}
[c]{|l|l|l|}\hline
$L$ & $\cdots$ & $L$\\\hline
\end{tabular}
}}  &  \otimes\underset{m_{2}+1}{\underbrace{%
\begin{tabular}
[c]{|l|l|l|}\hline
$L$ & $\cdots$ & $L$\\\hline
\end{tabular}
}}=0. \label{aa}%
\end{align}
Similarly, we have in equation (\ref{aa}),

\begin{enumerate}
\item for $l_{1}=0$,%
\begin{equation}%
\begin{tabular}
[c]{|l|l|l|}\hline
$T$ & $\cdots$ & $T$\\\hline
\end{tabular}
\underset{1}{\underbrace{%
\begin{tabular}
[c]{|l|l|l|}\hline
$L$ & $\cdots$ & $L$\\\hline
\end{tabular}
}}\otimes%
\begin{tabular}
[c]{|l|l|l|}\hline
$L$ & $\cdots$ & $L$\\\hline
\end{tabular}
=0,
\end{equation}

\item if for $l_{1}=2m$,%
\begin{equation}%
\begin{tabular}
[c]{|l|l|l|}\hline
$T$ & $\cdots$ & $T$\\\hline
\end{tabular}
\ \ \underset{2m-1}{\underbrace{%
\begin{tabular}
[c]{|l|l|l|}\hline
$L$ & $\cdots$ & $L$\\\hline
\end{tabular}
\ \ }}\otimes%
\begin{tabular}
[c]{|l|l|l|}\hline
$L$ & $\cdots$ & $L$\\\hline
\end{tabular}
\ \ =0,
\end{equation}
then (\ref{aa}) implies%
\begin{equation}%
\begin{tabular}
[c]{|l|l|l|}\hline
$T$ & $\cdots$ & $T$\\\hline
\end{tabular}
\ \ \underset{2m+1}{\underbrace{%
\begin{tabular}
[c]{|l|l|l|}\hline
$L$ & $\cdots$ & $L$\\\hline
\end{tabular}
\ \ }}\otimes%
\begin{tabular}
[c]{|l|l|l|}\hline
$L$ & $\cdots$ & $L$\\\hline
\end{tabular}
\ \ =0.
\end{equation}

\end{enumerate}
\end{proof}

\end{document}